\UseRawInputEncoding
\documentclass[showpacs,twocolumn,superscriptaddress,nofootinbib]{revtex4-2} 

\usepackage{xcolor, multirow}
\usepackage{hyperref}
\hypersetup{colorlinks=true,linkcolor=blue,citecolor=blue}
\usepackage{tikz}
\usepackage{graphicx}
\usepackage{xcolor, soul}
\sethlcolor{green}
\usepackage{dcolumn}
\usepackage{bm}
\usepackage{color}
\usepackage{enumitem}
\usepackage{mathrsfs}
\usepackage{amsmath}
\usepackage{amssymb}
\usepackage{cleveref}
\usepackage{orcidlink}

\crefname{equation}{Eqn.}{Eqns.}
\crefname{figure}{Fig.}{Figs.}
\crefname{section}{Sec.}{Sec.}
\crefname{table}{Table}{Tables}
\usepackage{physics}

\providecommand{\dif}{\mathrm{d}} \def\d{\dif}
\setstcolor{red}

\newcommand{\beq}{\begin{equation}}
\newcommand{\eeq}{\end{equation}}
\newcommand{\bea}{\begin{eqnarray}}
\newcommand{\eea}{\end{eqnarray}}
\newcommand{\non}{\nonumber}
\providecommand{\dif}{\mathrm{d}} 

\graphicspath{{pic/}}

\begin{document}

\title{Structure-preserving numerical simulations of test particle dynamics around slowly rotating neutron stars within Hartle-Thorne approach}

\author{Misbah Shahzadi
\orcidlink{0000-0002-3130-1602}}
\email{misbahshahzadi51@gmail.com}
\affiliation{Department of Mathematics, COMSATS University Islamabad, Lahore Campus, 54000 Lahore, Pakistan}

\author{Martin Kolo{\v s}
\orcidlink{0000-0002-4900-5537}}
\email{martin.kolos@physics.slu.cz}
\affiliation{Research Centre for Theoretical Physics and Astrophysics, Institute of Physics, \\Silesian University in Opava, Bezru\v{c}ovo n\'{a}m.13, CZ-74601 Opava, Czech Republic}

\author{Rabia Saleem
\orcidlink{0000-0002-1124-9624}}
\email{rabiasaleem@cuilahore.edu.pk}
\affiliation{Department of Mathematics, COMSATS University Islamabad, Lahore Campus, 54000 Lahore, Pakistan}

\author{Yousaf Habib
\orcidlink{0000-0003-1709-1159}}
\email{yousaf.habib@numl.edu.pk}
\affiliation{National University of Modern Languages, Lahore Campus, Lahore, Pakistan}
\affiliation{Department of Mathematics, University of California, San Diego, US}

\author{\\ Adri\'an Eduarte-Rojas
\orcidlink{0000-0002-1917-5666}}
\email{adrian.eduarte@ucr.ac.cr}
\affiliation{Space Research Center (CINESPA), School of Physics, University of Costa Rica, 11501-2060 San Jos\'e, Costa Rica}

\date{\today}
\begin{abstract}
In this paper, we explore the chaotic signatures of the geodesic dynamics for particles moving in the slowly rotating Hartle-Thorne spacetime; an approximate solution of vacuum Einstein field equations describing the exterior of a massive, deformed, and slowly rotating compact object. We employ the numerical study to examine the geodesics of prolate and oblate deformations for generic orbits and find the plateaus of the rotation curve, which are associated with the existence of Birkhoff islands in the Poincar\'{e} surface of the section, where the ratio of the radial and polar frequency of geodesics remains constant throughout the island. We investigate various phase space structures, including hyperbolic points and chaotic regions in the neighborhood of resonant islands. Moreover, chaotic behavior is observed to be governed by the stickiness phenomenon, where chaotic orbits remain attached to stable ones for an extended duration before eventually diverging and are attracted toward the surface of the neutron star. The precision of the numerical integration used to simulate the particle's trajectories plays a crucial role in the structures of the Poincar\'{e} surface of the section. We present a comparison of several efficient structure-preserving numerical schemes of order four applied to the considered non-integrable dynamical system and we investigate which schemes possess the canonical property of the Hamiltonian flow. In particular, we compare the performance of the symplectic Runge--Kutta integrator with the G-symplectic general linear method. Among the class of non-symplectic integrators, we employ the explicit Runge--Kutta method and explicit general linear method with a standard projection technique to project the numerical solution onto the desired manifold. The projection scheme admits the integration without any drift from the desired manifold and is computationally cost-effective. We are concerned with two crucial aspects -- long-term behaviour and CPU time consumption.
\end{abstract}
\maketitle

\setcounter{tocdepth}{2}
\tableofcontents

\section{Introduction} \label{intro}


Compact objects such as neutron stars and black holes (BHs) are ideal astrophysical laboratories to test the strong gravitational field regime predicted by general relativity. There are two classes of spacetime that may represent the exterior of compact objects: those that model spherical objects such as Schwarzschild and Kerr metrics, and those that depict the deformed compact objects, like the Hartle-Thorne (HT) metric. The Schwarzschild spacetime represents either non-rotating BH or the exterior of the non-rotating star. In contrast, the Kerr metric describes a rotating star, but only to linear order in the star's angular velocity $\Omega$. At higher orders, the multipole moments of the gravitational field created by a rapidly rotating compact star differ from those of a BH. This difference in the multipolar structure has important consequences for the observation of the gravitational and electromagnetic radiations from these objects. In principle, the gravitational waves emitted by particles around compact stars or BHs can be used to map the multipolar structure of the corresponding spacetime and check the validity of the no-hair theorem \cite{Ryan:1995:PhRvD:,Col-Hug:2004:PhRvD:}. 

The HT spacetime, introduced by Hartle and Thorne in 1968 \cite{Har-Tho:1968:ApJ:}, describes the structure of vacuum spacetime in the vicinity of slowly rotating neutron stars, constructed as a perturbation of a corresponding spherically symmetric non-rotating solution, with the perturbation being taken up to second order in the star's angular velocity $\Omega$. In this approximation, the spacetime describing the exterior of a slowly rotating neutron star is fully described by three parameters (mass, spin, and quadrupole), with the inner boundary of the exterior region being given by the stellar radius $R$. An invariant comparison of three different models of neutron stars (HT slow-rotation approximation, the exact analytic vacuum solution of Manko, and the numerical solution of the Einstein equations) shows that the HT approximation is very reliable for most astrophysical applications \cite{Berti-etal:2005:MNRAS:}. BHs are considered to be perfect spheres, while other compact objects like neutron stars may be slightly deformed by their rotation, hence the HT metric can be used to model neutron star exteriors where the Kerr metric may no longer be valid. 

The HT spacetime incorporates the Lense-Thirring effect \cite{Vieira-etal:2022:PhRvD:}, demonstrates frame dragging, and gives rise to significant precession effects \cite{Abramowicz-etal:2003:} and thus it has been particularly employed to deduce several astrophysical phenomena. Many authors have extensively investigated the HT spacetime in literature, exploring various phenomena such as epicyclic oscillations \cite{Urbancov-etal:2019:ApJ:}, quasinormal modes \cite{All-etal:2019:PhRvD:}, photon orbits and shadow \cite{Kos-Pap:2022:CQGra:}. Additionally, a ray-tracing algorithm to compute the apparent surface areas of moderately spinning neutron stars making use of the HT spacetime has been developed \cite{Michi-etal:2012:ApJ:}. The non-integrability of HT spacetime with prolate deformations for photon orbits \cite{Kos-Pap:2022:CQGra:} and massive particle orbits \cite{Kyriakos-Kostas:2023:arXiv230518522D:} has been discussed very recently.

The theoretical study of test particle non-linear dynamics in HT spacetime and the existence of strong $2:3$ resonance \cite{Kyriakos-Kostas:2023:arXiv230518522D:}, opens an interesting application as an explanation of quasi-periodic oscillation (QPO) observed in X-ray signal coming form various compact sources. The origin of QPOs is not yet fully understood, and there are several theoretical models proposed to explain them, but some connection to test particle orbital motion is eminent since the frequencies of the epicyclic motion around the central compact object are comparable to observed QPOs frequencies. In the low-mass X-ray binary systems containing neutron stars the frequencies of observed QPOs range from $\sim 10^{-2}$~Hz up to $\sim 10^3$~Hz. The high-frequency QPOs, with frequencies in the range $200-1300$~Hz are comparable to the frequencies of the orbital motion in strong gravity near neutron stars \citep{vdK:2006:CompStell:}. High frequency QPOs for the neutron star sources are usually observed as simultaneous peaks in the X-ray flux and when two frequencies are detected, they occur with a fixed small-number ratio, typically $3:2$ ($2:3$)\citep{Tor-etal:2005:ASTRA:}. The observed twin HF QPOs span a large frequency range following an approximately linear relation between both QPOs peeks \citep{Bel-etal:2005:ASTRA:}. The frequency ratio in twin HF QPOs changes in the range of $3:2$ to $5:4$ but a mostly resonant ratio $3:2$ is observed. Test particle non-linear dynamics in HT metric could play a crucial role for resonance emergence in the accretion discs oscillation modes around neutron stars, especially for $3:2$ and other small ratio resonances.

Numerical integration schemes are powerful tools to explore the behaviour of various non-linear dynamical systems in theoretical astrophysics; for example in the motion of relativistic test particles with the presence or absence of external electromagnetic fields \cite{Kop-Kar:2014:ApJ:,Kol-Stu-Tur:2015:CQGra:,Sharif-Shahzadi:2017:EPJC:,Pan-Kol-Stu:2019:EPJC:,Martin:2023:EPJC:}, as well as general relativistic systems involving compact binaries consisting of BHs or neutron stars \cite{Har-Tho:1968:ApJ:}. These schemes can be classified into two main types: explicit and implicit methods \cite{Harier:2000:Book}. Explicit methods compute the solution at a given time step based solely on the previous time step, while implicit methods involve solving an equation that depends on both the previous and current time steps. Numerical methods for ordinary differential equations (ODEs) can further be categorized as one-step methods, multi-step methods, and general linear methods (GLMs) \cite{Eirola-etal:1992:conservation:,Butcher:2006:GLM:,glm_underlying_onestep_method}.

The class of problem we are dealing with in this paper falls in the Hamiltonian systems, which possess various conserved quantities, including symplecticity of the flow, energy, and angular momenta, usually known as the first integrals since they remain constant throughout the system evolution. Symplectic numerical methods play a crucial role in accurately simulating the dynamics of Hamiltonian systems, preserving their inherent properties such as symplecticity and first integrals \cite{Harier:2000:Book,Channell:1990:,Sanz-Jesus:1992:,SRK_Sanz,Sanz-Serna:2018:numerical:}. These methods, including symplectic Runge-Kutta (RK) and GLMs, aim to maintain the symplectic structure of the system, ensuring that the numerical solutions accurately capture the conservative nature of Hamiltonian dynamics, preserving energy and phase space properties \cite{Butcher:2006:GLM:,Butcher:2016:BIT:}. The projection technique \cite{Hairer:2000:symmetric:} can also be applied to enhance the numerical solution of ODEs with known invariants to stay on a manifold described by the invariants. The advantage of this approach is the ability to use explicit methods with projection techniques, which can conserve the invariants while having lower computational costs.

In recent times, there has been a growing interest among researchers in the development of symplectic integrators for Hamiltonian systems. These integrators are then tested using BH solutions immersed in a uniform magnetic field \cite{Wang-etal:2021:Schw,Wang-etal:2021:RN,Wu-etal:2021:Kerr,Zhou-etal:2022:}. The numerical integrators have also been introduced and tested for massive particles in non-standard spacetimes and post-Newtonian systems in \cite{Seyrich-Gera:2012:PhRvD:,Seyrich:2013:PhRvD:,Witzany-etal:2015:MNRAS:,Hainzl-Seyrich:2016:EPJB:,Witzany:2018:MNRAS:} A comparison of explicit methods with the implicit solution of the discretized equations of motion for charged particles in electromagnetic fields is presented in \cite{Ripperda-etal:2018:ApJS:}. The performance of the implicit midpoint scheme against the standard fourth-order RK explicit integrators has been tested on a number of standard and non-standard spacetimes, using simulations of both photons and massive particle trajectories \cite{Kopack:2014:ragt.conf:,Bacchini-etal:2018:,Bacchini-etal:2019:}.

The intention of this work is to explore the chaotic dynamics in the background of a slowly rotating neutron star based on HT formalism. We explore several structures in phase space such as Birkhof chains of islands, chaotic regions, hyperbolic points, and higher-ordered islands by anlysing the Poincar\'{e} surface of sections (PSs) and rotation curves. We show that most of the chaotic orbits are sticky chaotic orbits that remain attached to the stable orbits for a long period of time. To check the accuracy of numerical integration used for simulating the time evolution of equations of motion, we present several efficient numerical integration techniques, including symplectic and non-symplectic schemes. In particular, we compare the performance of the symplectic RK integrator with the G-symplectic GLM. Within the class of non-symplectic integrators, we use explicit RK integrators and explicit GLMs, employing a standard projection technique to project the numerical solution onto the desired manifold. We compare the performance of numerical integrators in preserving the qualitative features of the system.

We use the spacelike signature $(-,+,+,+)$, geometric units $G=c=1$, and Greek indices are defined to range from $0$ to $3$. 

\section{Chaotic dynamics in HT spacetime} 

In this section, we explore the dynamical features of the system consisting of HT spacetime.

\subsection{The HT spacetime} \label{HT}

The HT metric is an approximate vacuum solution of Einstein field equations that describes the exterior of a slowly rotating deformed compact object and constructed perturbative in terms of the rotation rate. It is fully characterized by the source mass $M$, angular momentum $J=a/M$ (up to the second order), and quadrupole moment $q$ (up to the first order), and given by \cite{Har-Tho:1968:ApJ:,Bos-etal:2012:PhRvD:}  
\bea\non
\d s^2 &=& - f_{1} \left[\{ 1 + 2 f_{2} P_{2}(\theta) \} - \frac{2 a^2}{r^4} (2 \cos^{2}\theta - 1) f_{1}^{-1} \right] \d{t}^2 \\\non
&+& f_{1}^{-1} \left[1 - 2 \left(f_{2} - \frac{6 a^2}{r^4} \right) P_{2}(\theta) - \frac{2 a^2}{r^4} f_{1}^{-1} \right] \d{r}^2 \\\non
&+& r^2 \left[1 - 2f_{3} P_{2}(\theta) \right] \left[\d\theta^2 + \sin^{2}\theta\, \d\phi^2 \right]\\\ 
&-& \frac{4a}{r} \sin^{2}\theta\, \d{t} \d \phi, \label{HT-Metric}
\eea
where $P_{2}(\theta) = (3 \cos^{2}\theta - 1)/2$ is the Legendre polynomial and $a$ denotes the spin parameter. The unknown functions $f_{1}$, $f_{2}$, and $f_{3}$ take the form
\bea
f_{1} &=& \left(1- \frac{2 M}{r} \right),\\\
f_{2} &=& \frac{a^2}{M r^3} \left(1 + \frac{M}{r}\right) + \frac{5}{8} \left(\frac{q}{M^3} - \frac{a^2}{M^4} \right)\, Q^{2}_{2} (x),\\\
f_{3} &=&  f_{2} + \frac{a^2}{r^4} + \frac{5}{4} \left(\frac{q}{M^2 r} - \frac{a^2}{M^3 r} \right)\, f_{1}^{-1}\, Q^{1}_{2} (x).
\eea
The associated Legendre functions $Q^{1}_{2} (x)$ $Q^{2}_{2} (x)$ of the second kind can be written as
\bea
Q^{1}_{2} (x) &=& \sqrt{x^2 - 1} \left[\frac{3}{2} x \ln(\frac{x + 1}{x - 1}) - \frac{3x^2 - 2}{x^2 - 1} \right],\\\
Q^{2}_{2} (x) &=& (x^2 - 1) \left[\frac{3}{2} \ln(\frac{x + 1}{x - 1}) - \frac{3x^3 - 5x}{(x^2 - 1)^2} \right],
\eea
where $x= (\frac{r}{M} - 1)$. The quadrupole mass moment $q$ measures the deviation from a spherical gravitational source. The HT spacetime (\ref{HT-Metric}) corresponds to an oblate object when the quadrupole moment is positive ($q>0$), and to a prolate object when it is negative ($q<0$). For $q=0=a$, the HT spacetime (\ref{HT-Metric}) leads to the Schwarzschild one. The Kerr spacetime with angular momentum up to the second order, using the Boyer-Lindquist coordinates, can be obtained from HT spacetime (\ref{HT-Metric}) by taking $q=J^2$ and using the coordinate transformations
\bea\non
r_{\rm BL} &=& r - \frac{a^2}{2 r^3} [(r + 2 M)(r - M) \\\ &+& (r - 2 M)(r + 3 M) \cos^{2} \theta],\\\
\theta_{\rm BL} &=& \theta - \frac{a^2}{2 r^3} (r + 2 M) \cos\theta \sin \theta.
\eea
The HT metric (\ref{HT-Metric}) can also be reduced to the Erez-Rosen metric by appropriate coordinate transformation \cite{Boshkayev-etal:2019:Symm:}. There are two different ways to proceed when working with HT spacetime. The first approach is to use the metric (\ref{HT-Metric}) ``as it is" (truncated at a given $\Omega$-order), without making any further approximations in the geodesic equations. The second way is to expand all equations to the same perturbative order as the metric \cite{Kos-Pap:2019:PhRvD:}. We will use the first approach in order to integrate the equations of motion. Furthermore, we will assume that the HT spacetime ends at the surface of the compact object. 

Another metric closely related to the HT metric is the Kerr-like metric with mass quadrupole moment, which is an approximate solution of Einstein field equations \cite{Frutos:2016:IJAA:}. This metric describes the geometry of the spacetime surrounding a spinning compact object, where the mass quadrupole moment represents the deformation from a spherical object. Similar to the HT metric, the Kerr-like metric also has three parameters to characterize the spacetime (mass, spin parameter, and mass quadrupole moment). The transformation between these two metrics using Taylor expansion is developed by Frutos \cite{Frutos}. The study of chaotic behaviour in the context of this non-Kerr metric has been explored in \cite{Adrian-etal:2022:PhRvD:}.

\subsubsection{Geodesic motion in HT spacetime} 

The motion of a neutral test particle with mass $m$ can be described by the Hamiltonian given by
\beq\label{Hamiltonian}
H=\frac{1}{2m}g^{\alpha \beta} p_{\alpha} p_{\beta},
\eeq
where $p^{\alpha}=m u^{\alpha}$ denotes the four-momentum, $u^{\alpha}=\frac{d x^{\alpha}}{d\tau}$ is the four-velocity, and $\tau$ is the proper time of the particle. Equations of motion of particles can be found using Hamiltonian formalism written in the form
\beq
\frac{dx^{\alpha}}{d\zeta} \equiv m u^{\alpha}=\frac{\partial H}{\partial p_{\alpha}}, \quad
\frac{d p_{\alpha}}{d\zeta} = -\frac{\partial H}{\partial x^{\alpha}},
\eeq
where $ \zeta=\tau/m$ is the affine parameter. Due to stationary and axisymmetric properties of the spacetime (\ref{HT-Metric}), the Hamiltonian (\ref{Hamiltonian}) is independent of the coordinates $t$ and $\phi$. Therefore, the momenta $p_{t}$ and $p_{\phi}$ corresponding to these coordinates remain constant during the motion. This indicates that both $p_{t}$ and $p_{\phi}$ are integrals of motion, which ensures that the specific energy $\mathcal{E}=E/m$ and specific angular momentum $\mathcal{L}=L/m$ stay conserved throughout the motion and can be expressed as
\bea\label{energy}
\frac{p_{t}}{m}&=&g_{tt}u^{t}+g_{t\phi}u^{\phi}=-\mathcal{E},\\
\frac{p_{\phi}}{m}&=&g_{\phi\phi}u^{\phi}+g_{t\phi}u^{t}=\mathcal{L}.\label{ang-mom}
\eea
Since the Hamiltonian system (\ref{Hamiltonian}) is an autonomous system ($\d H/ \d \tau = \partial H/\partial \tau =0$), thus $H$ itself is a third integral of motion, given by $H=-m^{2}/2$, because $g^{\alpha\beta} p_{\alpha} p_{\beta} =-m^2 $. The first two integrals of motion ((\ref{energy})-(\ref{ang-mom})) can be used to reduce the number of degrees of freedom from four to two. Thus, the motion can be restricted to the meridian plane, i.e., the $r-\theta$ plane. 

The phase space of test particle dynamics in the Kerr spacetime contains only the main island of stability as a consequence of its integrability, with no other structures present. In contrast to the Kerr metric, the HT metric does not possess the Carter constant, and such symmetry is perturbed by the presence of quadrupole parameter $q$. HT metric is a non-Kerr rotating spacetime with higher-order moments (quadrupole) deviating from a Kerr spacetime, HT geodesics are non-integrable and dynamics of test particles can exhibit chaotic behaviour.

\subsection{Non-linear dynamics}\label{chaos}

When dynamical systems are subject to non-integrable perturbations, they display special characteristics in contrast to their unperturbed counterparts. One significant effect is the emergence of chaotic motions in specific areas of the phase space, triggered by the perturbation. Deterministic chaos may completely dominate the dynamics, depending on the strength of the perturbation. However, even in the slightly perturbed systems with negligible chaos, one can still observe significant non-integrable effects in the vicinity of resonances. 

The transition from integrability to non-integrability is governed by two fundamental theorems, namely Kolmogorov-Arnold-Moser (KAM) theorem \cite{Goldstein:2002:book:} and Poincar\'{e}-Birkhoff theorem \cite{Birkhoff:1913:}. Both theorems play an important role in Hamiltonian systems perturbed by small parameters. In a Hamiltonian system consisting of N oscillators, there exists a 2N-dimensional phase space where an N-dimensional torus is embedded, restricting the movement of a test particle to this torus. 

According to the KAM theorem, if the bounded motion of an integrable Hamiltonian system $\mathcal{H}_{0}$ is perturbed by a small $\Delta \mathcal{H}_{0}$ such that the total Hamiltonian $\mathcal{H} = \mathcal{H}_{0} + \Delta \mathcal{H}_{0}$ becomes non-integrable, for small perturbations, most of the non-resonant invariant tori that are far enough from resonances are deformed, but they are not destroyed. The newly deformed tori are called KAM tori, and the corresponding PSs closely resemble that of the corresponding integrable system. However, the dynamics of the system in the vicinity of resonances can vary significantly, which may affect the measurable properties of the system. 

The Poincar\'{e}-Birkhoff theorem states that when a system undergoes a small deviation from an integrable system, the resonant invariant curve breaks apart, and only a finite even number ($2km$, where $k\in N$) of the periodic points of the period $m$ survive. This means that only $2k$ periodic orbits remain from the resonant torus, half of which are stable and the other half are unstable. By visualizing a closed curve that passes through all surviving points of the disintegrated resonant curve, the stable and unstable periodic points alternate along the curve, creating the Birkhoff chain. Each stable periodic point is surrounded by a set of nested KAM curves, forming an island of stability. In a resonant case, the phase-orbit visits all $m$ islands of the n/m-resonance by moving sequentially to the next n-th island along the aforementioned closed curve. At every step, it winds and eventually creates the KAM curves inside each island.

\subsubsection{Signatures of chaos}

\begin{figure}[h!]
\centering
\includegraphics[width=\hsize]{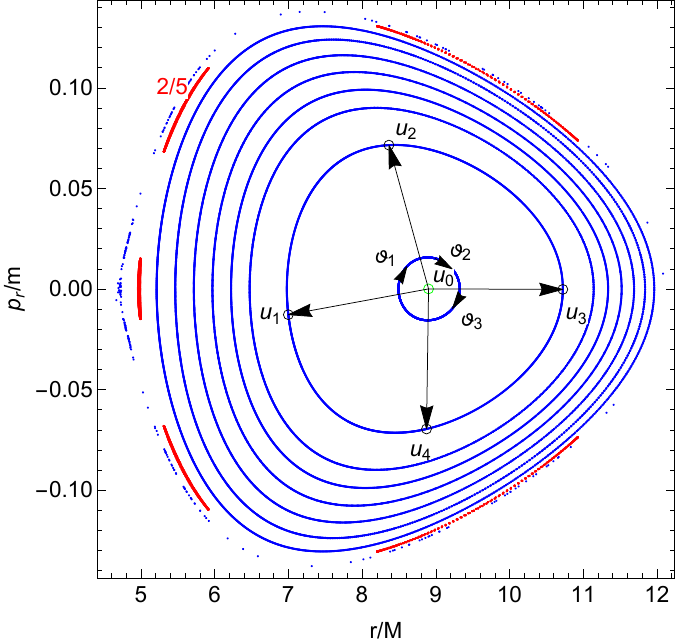}
\caption{PSs for $\theta=\pi/2$, $p_{r}/m = 0$, $p_{\theta}>0$, and parameters $q/M^3=1.5$, $\mathcal{E}=0.95$, $\mathcal{L}/M=3$, and $J/M^2=0.5$. The central part of the figure represents the procedure to compute the rotation number. The green circle $\textbf{u}_{0}$ indicates the position of the invariant point, while the black arrows point towards the position of the crossing in the PS, denoted with black circles. The PS in red color corresponds to the $2/5$-resonance. 
\label{intro_fig}}
\end{figure}

In order to gain better insight into orbital phenomena and chaotic imprints, well-established tools can be utilized to analyze the phase space structure of orbits. A widely used technique is the PS, which is a lower-dimensional subspace of phase space in a dynamical system, constructed by successive intersections of geodesics with a chosen 2-dimensional slice of the torus. Whenever the orbit intersects the slice, it generates a single point on that slice. The complete PS is produced by a sufficient number of successive intersections, with strictly positive or negative directions of the intersection, as shown in Fig.~\ref{intro_fig}. The presence of chaos can be instantly revealed by analyzing the structure of the PS.

The regular dynamics is characterized by periodic and quasi-periodic orbits located on invariant tori, while the chaotic orbits fill the available domain in phase space densely. If the perturbation parameter of a 2-dimensional non-integrable system is small enough, the Birkhoff islands of stability are extremely thin and their detection on the surface of the section is very difficult. Thus, one needs to find suitable initial conditions of orbits that will eventually result in a chain of islands on the surface of the section. However, a more advanced method can be used to detect these islands. The islands of stability lie around a resonant periodic orbit which is defined by a commensurate ratio of frequencies $\nu_{\vartheta} = \omega_{r}/\omega_{\theta} = n/m$, where $\omega_{r}$ and $\omega_{\theta}$ are radial and vertical frequencies, and $n,m$ are integers. This frequency ratio identifies not only the resonant periodic orbit of the island, which is a collection of $m$ stable points on a surface of the section but also all the KAM orbits belonging to the particular chain of islands around them. Although each distinct KAM orbit on an island is distinguished by a different pair of frequencies $\omega_{r}$ and $\omega_{\theta}$, all of them share the same commensurate ratio $\nu_{\vartheta}$ with the central periodic orbit \cite{Gera-etal:2010:PhRvD:,Gerakopoulos:2012:PhRvD:}.

The existence of KAM invariant tori within regular islands in two-dimensional area-preserving maps results in the division of phase space into separate regions. Within these regions, orbits in the chaotic sea will never enter any island, and periodic and quasi-periodic orbits residing within an island will never transition to the chaotic sea \cite{Lichtenberg:2013:,Mackay-etal:1984:PhyD:}. The presence of embedded islands within the chaotic sea constitutes a fractal structure, and it is challenging to determine exactly the island's boundary \cite{Umberger:1985:prl}. The islands are surrounded by smaller islands, which are in turn surrounded by higher-order islands that eventually form a chaotic layer. The hierarchical islands-around-islands structure continues infinitely and repeats itself at any arbitrarily small scale \cite{Meiss-Ott:1986:PhyD:}.

In a non-integrable dynamical system, both regular and chaotic trajectories may coexist in the phase space. The standard method for qualitatively investigating non-linear dynamics involves constructing PS which allows us to visually discriminate between the chaotic and regular regime of motion. In certain weakly chaotic dynamical systems, the PS seems to be quite regular, with no prominent indication of chaos. However, a detailed examination of the structures of PSs is crucial for uncovering signs of chaos. Despite the fact that the PSs display all the dynamical features, such as Birkhoff chains of islands or deformed tori, these regions can be extremely small in phase space and difficult to observe. Thus, detecting these structures directly in the PS would be challenging since the readability of the PS decreases rapidly with the increasing density of depicted trajectories. However, there is a powerful tool, the rotation number, that enables us to quantitatively analyse the properties of chaos. The rotation number corresponds to the PS and computes the ratio of fundamental frequencies. The classification of orbits can be determined using the rotation number. Rational rotation numbers indicate periodic orbits, which form closed curves on the torus and are known as resonant orbits. In contrast, irrational rotation numbers suggest quasi-periodic orbits that densely cover the torus. Most importantly, the rotation numbers can also identify the signatures of chaos, even if the chaotic behaviours are very weak \cite{Gera-etal:2010:PhRvD:}. In order to calculate the rotation number, first we identify the central invariant point $\textbf{u}_{0}$ of the PS. This is the fixed point corresponding to the periodic orbit which crosses the equatorial plane at only one point with $p_{r}/m = 0$ moving towards the positive part of the z-axis, see the green circle in Fig.~\ref{intro_fig}. Then, we define the position vector $\textbf{r}_{i}$ of the $i-th$ crossing point $\textbf{u}_{i}$ of a phase orbit on a PS as
\beq
\textbf{r}_{i} = \textbf{u}_{i} - \textbf{u}_{0},
\eeq
indicating its position relative to $\textbf{u}_{0}$. Using these vectors, we compute the so-called rotation angle that is subtended clockwise by them, given by
\beq\label{angle}
\vartheta_{i} = \angle(\textbf{r}_{i+1}, \textbf{r}_{i}),
\eeq
as shown in Fig.~\ref{intro_fig} for four points $\textbf{u}_{i}$ that belong to the same PS. This angle (\ref{angle}) is computed for each consecutive pair of piercings. Summing up all the angles $\vartheta_{i}$ and divided by $2 \pi n$, where $n$ represents the number of piercings that occurred in the corresponding PS, we obtain a so-called rotation number, given by 
\beq
\nu_{\vartheta} = \lim_{n \to \infty} \frac{1}{2 \pi n } \sum_{i=1}^{n} \vartheta_{i},
\eeq
which measures the average fraction of the orbit \cite{Contopoulos:2002:book:}. The rotation number usually appears to grow monotonically as long as we cross KAM curves that encircle the central fixed point $\textbf{u}_{0}$. On a resonant island, the rotation number remains fixed at a constant rational value, which is characteristic of the corresponding island of stability. For the chaotic regions of the non-integrable system, the rotation number exhibits irregular fluctuations from one point to another. Therefore, its behaviour appears to be smooth only in the area occupied by regular orbits. In a slightly perturbed integrable system, the chaotic layers surrounding the Birkhoff chains of islands are extremely thin, making the fluctuating behaviour of the rotation number difficult to observe \cite{Gera-etal:2010:PhRvD:,Alejandro-etal:2018:CQGra:,Adrian-etal:2022:PhRvD:}.

The usefulness of the rotation number goes beyond its ability to identify the dynamics of a system. It serves as an indicator of various phenomena. For instance, a plateau in the rotation number signals the presence of a constant ratio of the orbital frequencies, which then translates into a constant pattern of frequencies in the emitted gravitational waves. An observation of such a constant pattern would constitute a clear signal of the presence of chaos, and therefore a novel test of general relativity and Kerr hypothesis \cite{Gera-etal:2010:PhRvD:,Gera-Kop:2018:IJMPD:,Kyriakos-etal:2020:PhRvD:}.

By plotting the rotation number as a function of the distance of initial conditions from the central periodic orbit $\textbf{u}_{0}$ of the main island of stability in a specific direction, we obtain the so-called rotation curve. In integrable systems such as the Schwarzschild or Kerr metric, this curve is a smooth and strictly monotonic function. When the system is perturbed, the smoothness of the rotation curve is disrupted, and it exhibits an approximate monotonic behaviour. Indeed, the rotation curve reveals clear indications of chaos.

The details of the PS of a non-integrable Hamiltonian system close to resonances are quite different from those of integrable systems. Resonant chains are in principle detectable in terms of spectral analysis of the observed electromagnetic and gravitational wave signals coming from systems like extreme mass ratio inspirals \cite{Gera-etal:2010:PhRvD:}. The presence of Birkhoff chains enables us to differentiate between a perturbed system and a regular one. Moreover, the position and the width of the chains also reflect other properties of the system.

\begin{figure}[h!]
\centering
\includegraphics[width=\hsize]{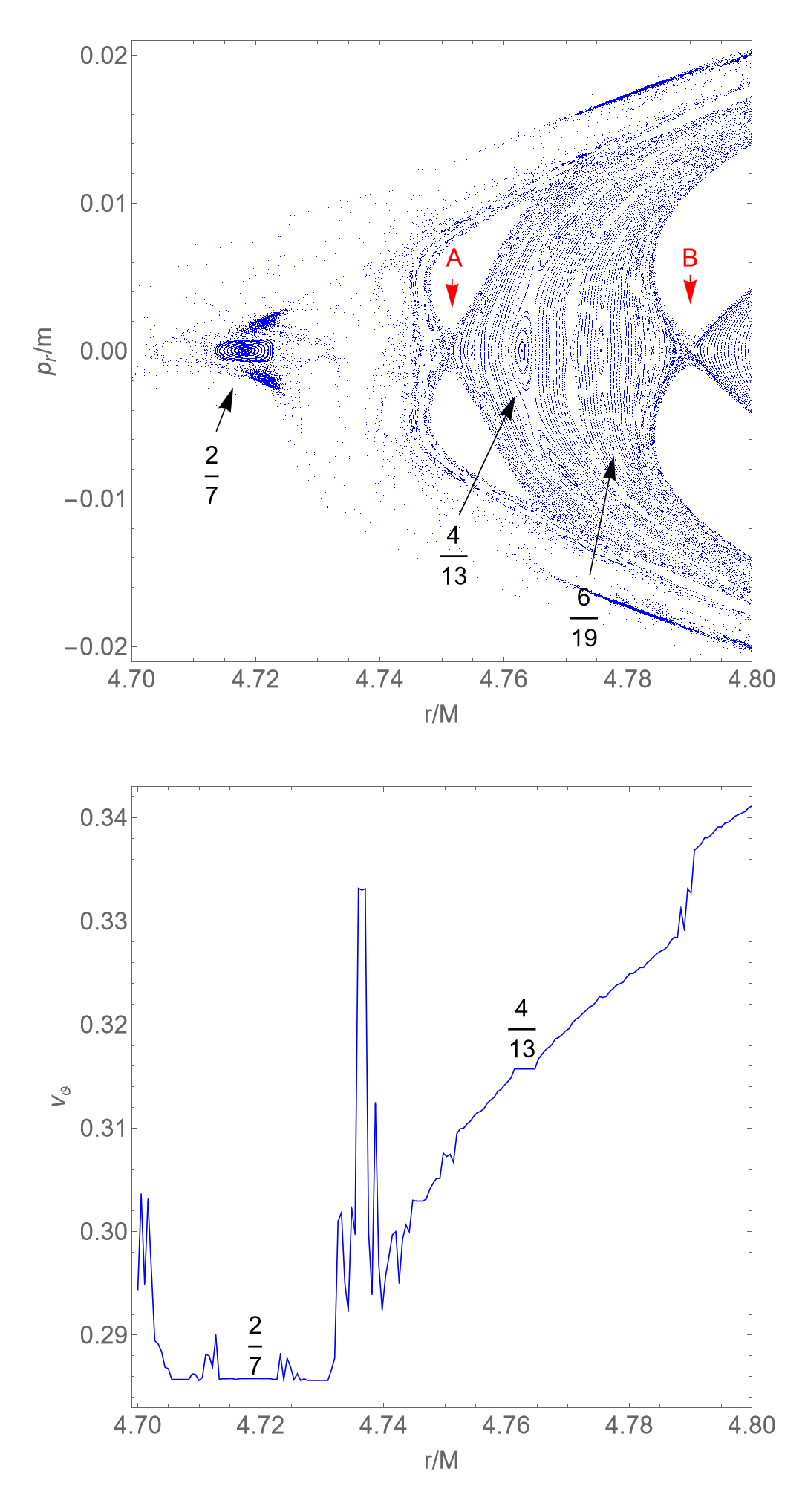}
\caption{Upper panel: a magnification of the left tip of the main island of stability presented in Fig.~\ref{intro_fig}. Various islands of stability forming Birkhoff chains are shown. Prominent islands of stability are labeled by the corresponding rotation numbers. The black arrows point toward the positions of islands of stability, while the red arrows (A and B) indicate the positions of the hyperbolic points along the line $p_{r}/m = 0$. Lower panel: rotation curve corresponding to the upper panel.
\label{fig2}}
\end{figure}

\begin{figure}[h!]
\centering
\includegraphics[width=\hsize]{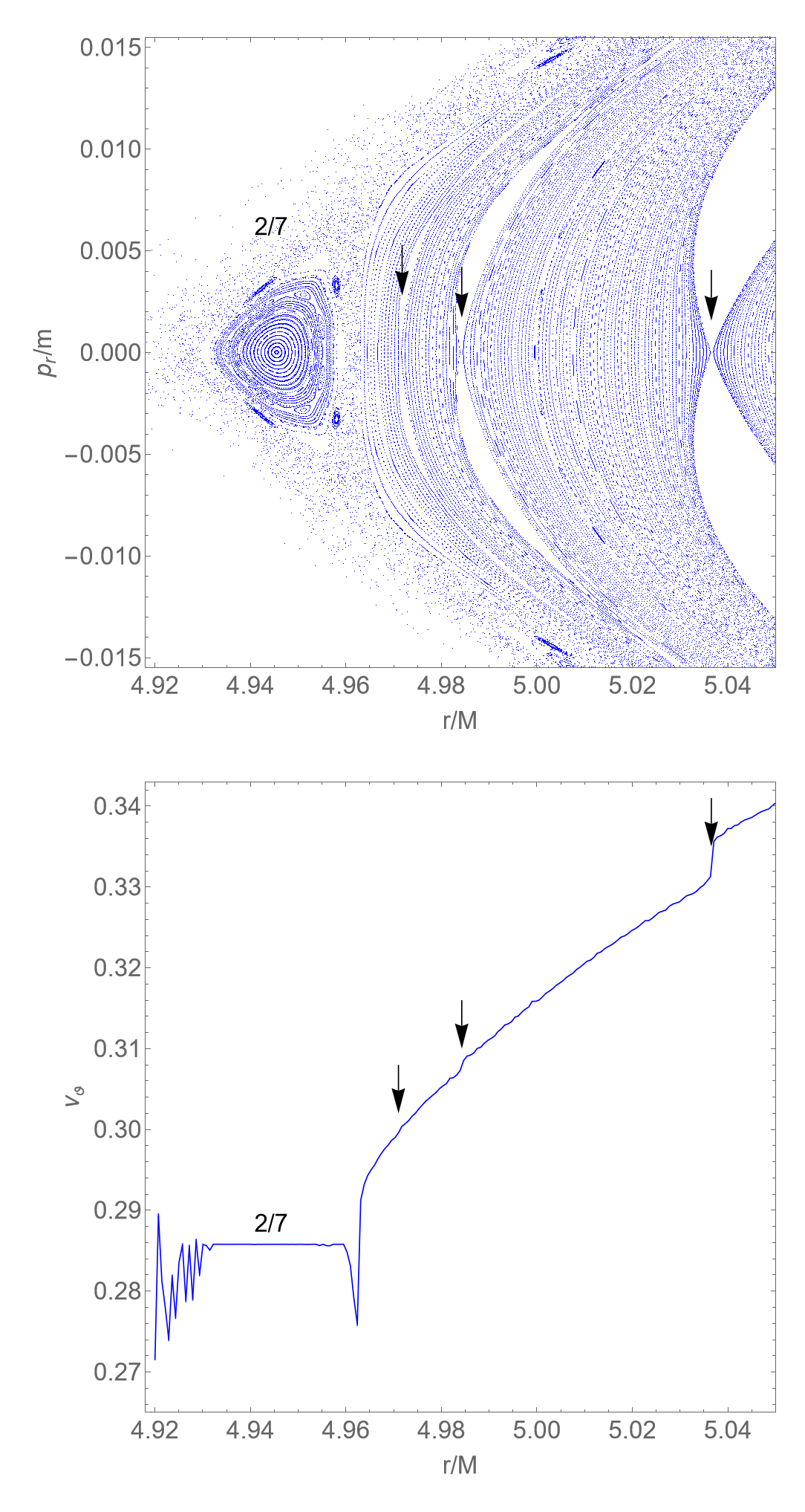}
\caption{Upper panel: a detail of the PS for $J/M^2=0.4$, $q/M^3=1.5$, $\mathcal{L}/M=3$, and $\mathcal{E} =0.95$. An island of stability embedded within a chaotic region, corresponding to the $2/7$ resonance is shown. The arrows represent the position of the unstable periodic points associated with other Birkhoff chains. Lower panel: rotation curve corresponding to the upper panel.
\label{fig4}}
\end{figure}

The PSs presented in Fig.~\ref{intro_fig} seem to be quite regular, filled densely with KAM curves, and display only one thin island of stability. There is no prominent indication of chaos. However, the existence of the Birkhoff chain, with multiplicity 5 (labeled as 2/5), implies that the system is indeed non-integrable and therefore chaos may be present. In fact, the PSs should be densely filled with other Birkhoff chains of islands as well, but their detection demands a very detailed scan of the PS. A detailed structure of the PS at the left tip of the main island of stability is presented in the upper panel of Fig.~\ref{fig2}, where chaotic behaviour can be observed. An island of stability associated with the $2/7$-resonant orbits, where $\nu_{\vartheta}=2/7$, is embedded in a sea of scattered points. The scattered points define the chaotic region, while the continuous curves define the limit of the regular domain. Moving along the $p_{r}/m = 0$ axis to the right, small islands of stability and hyperbolic points separating regions are identified. It can be seen clearly that the broken tori form the small chains of Birkhoff islands, the hyperbolic point $B$ separates the chaotic and regular regions, and the hyperbolic point $A$ is in the highly chaotic zone. The islands with lower multiplicity are more prominent than the islands with higher multiplicity. The multiplicity corresponds to the denominator of the prime number ratio associated with the rotation number. The large white spaces above and below the hyperbolic points in the PSs are embedded islands orbiting the main one.

The stickiness phenomenon, first reported by Contopoulos \cite{Contopoulos:2002:book:}, is a fundamental feature of the quasi-integrable Hamiltonian system that emerges from the coexistence of regular and chaotic regions. This coexistence results in the formation of regions that act like fractal scattering zones, situated near the boundaries of islands where chaotic trajectories are compelled to exhibit regular behaviour. Chaotic trajectories spend a long time in the vicinity of the island after crossing the barrier of a non-hyperbolic region, see upper panel of Fig.~\ref{fig2}. In order to better analyse the chaotic features, the rotation curve along the line $p_{r}/m = 0$ of the upper panel of Fig.~\ref{fig2} is shown in the lower panel of the same figure. The regions dominated by regular motion are depicted as relatively smooth segments along the curve, whereas the chaotic regions are identifiable by the presence of distinct portions where the rotation curve exhibits fluctuations. In specific smooth segments of the rotation curve, the rotation number remains constant. These segments resemble ``plateaus" within the rotation curve. Distinct plateaus in the lower panel correspond to the islands of stability associated with distinct resonances in the upper panel. The most prominent plateaus are labeled and identified by their corresponding rotation number values. The rotation curve exhibits a  relatively prominent jump at $r/M \sim 4.795$ indicating that the empty spaces above and below the hyperbolic point B correspond to the $1/3$ resonances. 

Another example for $q/M^3 >0$ showing the structure of the phase space is presented in Fig.~\ref{fig4} and the corresponding rotation curve is depicted in the bottom panel. The large island of stability corresponding to $2/7$-resonance is embedded in a chaotic layer. After the scattered points on the right of $2/7$-resonance, there is a region of regular orbits with small islands of stability and unstable periodic points, see black arrows in the upper panel. The irregular variations observed in the rotation number show the chaoticity of the orbits that surround the main island. The rotation curve shows a plateau belonging to a $2/7$ resonant island between irregular variations and changes abruptly when crossing the unstable periodic points, see black arrows. After the fluctuations on the right of the $2/7$ resonant island, the rotation curve grows like a strictly monotonic function. The rotation curve has a  relatively prominent jump at $r/M \sim 5.04$ which indicates that the empty spaces above and below the corresponding hyperbolic point correspond to the $1/3$ resonances. 

\begin{figure}[h!]
\centering
\includegraphics[width=\hsize]{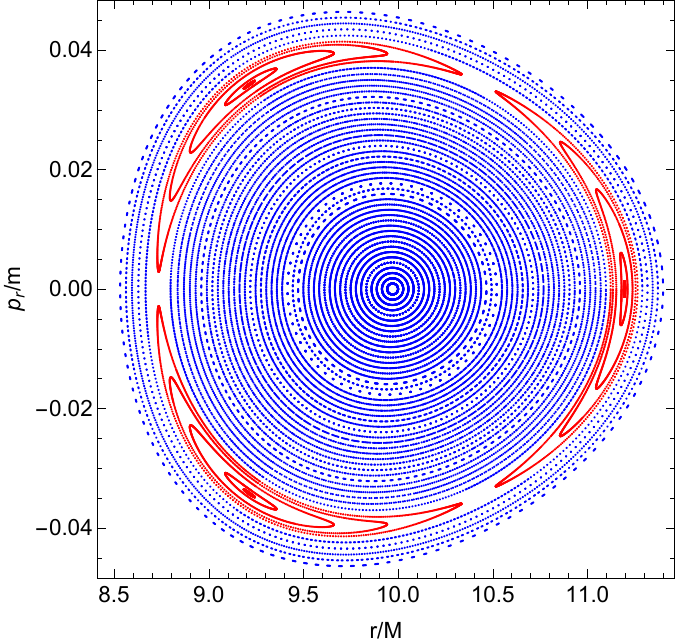}
\caption{PSs for $\theta = \pi/2$, $p_\theta >0$, and parameters $\mathcal{E} =0.955$, $\mathcal{L}/M=0.85$, $q/M^3=1$, $J/M^2=0.5$ along the axis $p_{r}/m = 0$. The PSs in red color correspond to $1/3$ resonances crossing $p_{r}/m = 0$.
\label{fig_RES}}
\end{figure}

The sizes and positions of islands depend on both the physical parameters of the metric (\ref{HT-Metric}) and the specific parameters of the orbit itself. Our analysis reveals that the most prominent resonant islands, characterized by their significant width, correspond to $1/3$ resonances, which can be clearly seen from Figs.~\ref{fig2} and \ref{fig4}, represented by the empty white spaces above and below the hyperbolic points. However, we can also observe it along the $p_{r}/m = 0$ axis at extended distances, such as $r/M > 10$, as illustrated in Fig.~\ref{fig_RES}.

\begin{figure}[h!]
\centering
\includegraphics[width=\hsize]{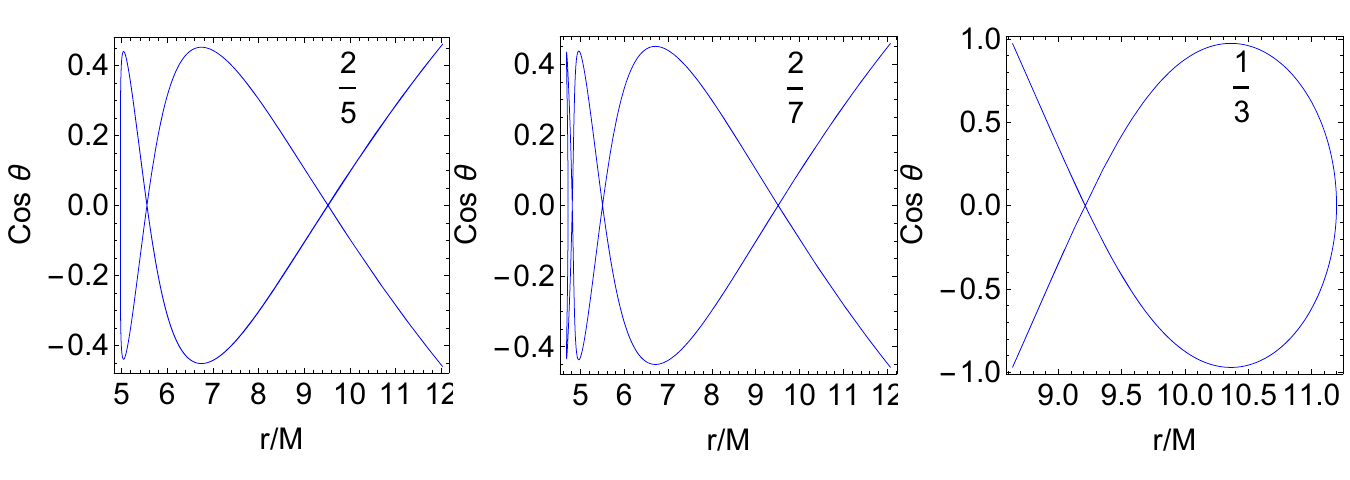}
\caption{Projection of the resonant orbits into the $r-cos\theta$ plane, corresponding to the orbits with resonances $2/5$, $2/7$ and $1/3$ presented in Figs.~\ref{intro_fig}, \ref{fig2} and \ref{fig_RES}, respectively. 
\label{rcosine}}
\end{figure}

\begin{figure}[h!]
\centering
\includegraphics[width=\hsize]{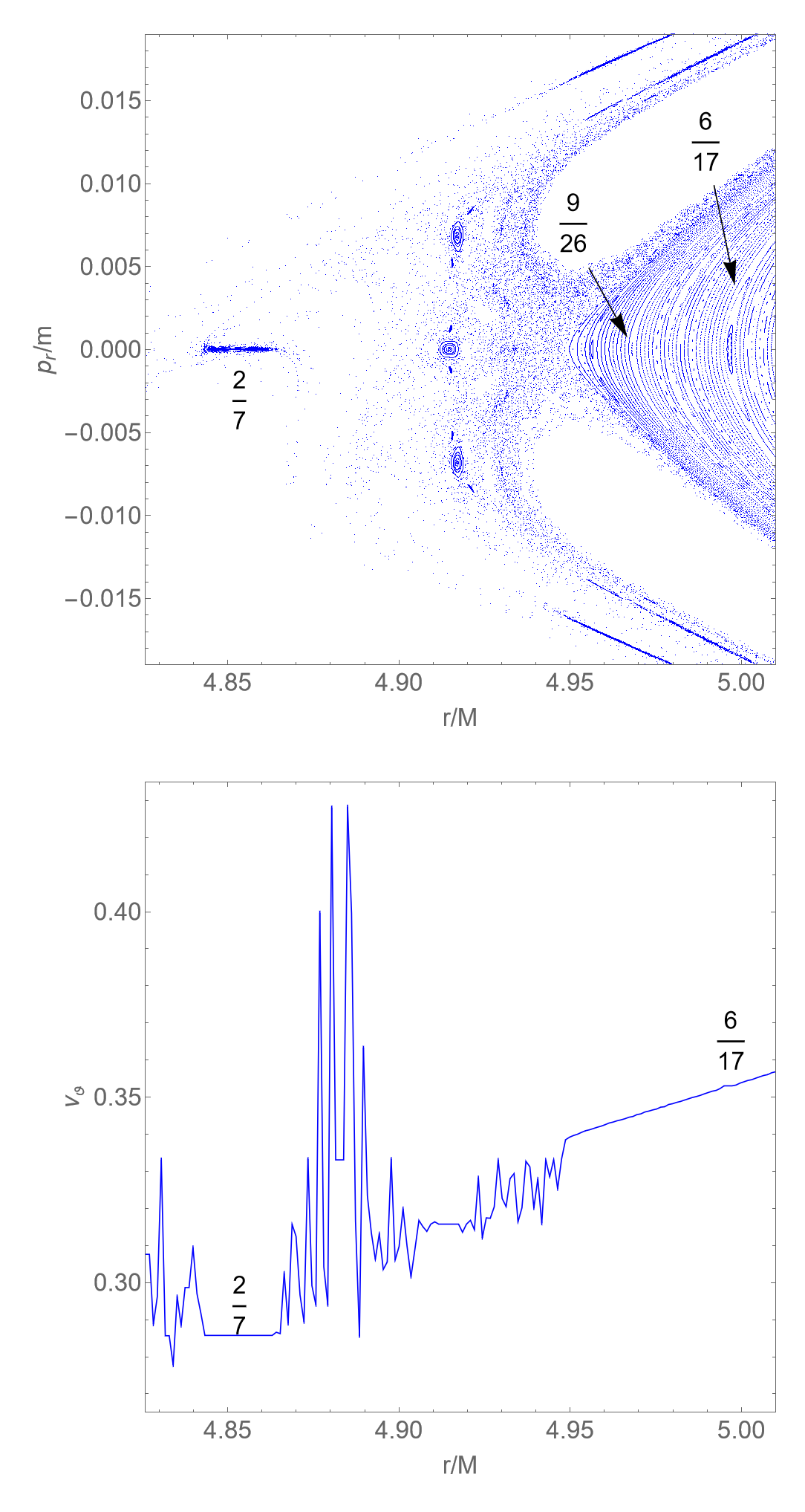}
\caption{Upper panel: sticky chaotic orbits surrounding the region of regular orbits for $J/M^2=0.4$, $q/M^3= -2$, $\mathcal{L}/M=1.15$, and $\mathcal{E}=0.954$. There are islands of stability of resonances $9/26$ and $6/17$ in the regular region. Several higher-ordered islands of stability are embedded within the sticky chaotic region. Lower panel: rotation curve corresponding to the upper panel.
\label{fig5}}
\end{figure}

The projection of three significant resonances, namely $2/5$, $2/7$, and $1/3$ as depicted in Figs.~\ref{intro_fig}, \ref{fig2} and \ref{fig_RES} respectively, onto the $r-\cos\theta$ plane is presented in Fig.~\ref{rcosine}. When the particles in resonances are projected into the $r-\cos\theta$ plane, they form closed curves that are related to the oscillations taking place in both the $r$ and $\theta$ coordinates. The resonant orbits no longer sample the available phase space, in contrast with generic orbits that cover it densely.

\begin{figure}[h!]
\centering
\includegraphics[width=\hsize]{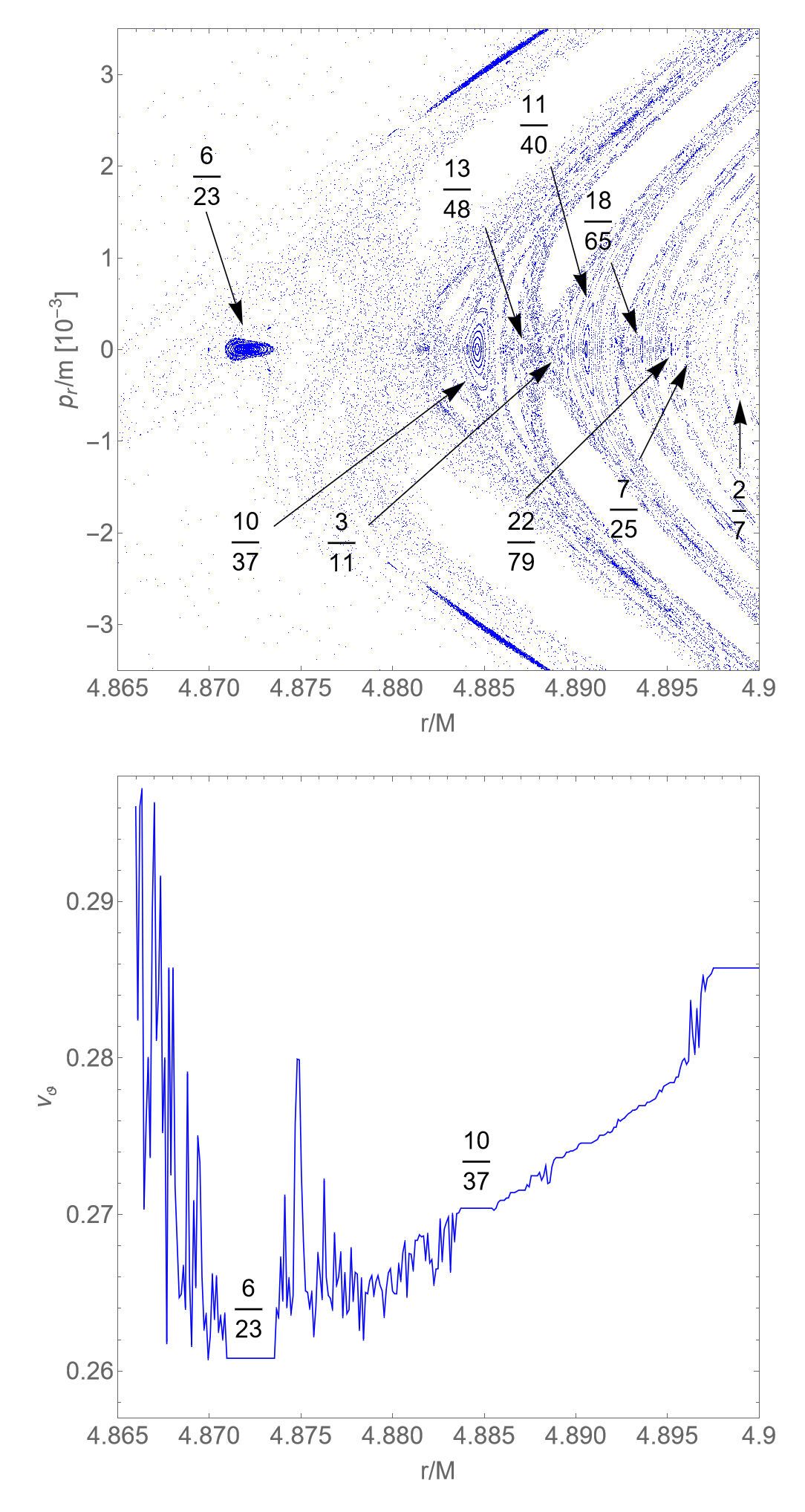}
\caption{Upper panel: a detail of the PS for parameters $\mathcal{L}/M=0.999$, $\mathcal{E}=0.95$, $J/M^2=0.4$, and $q/M^3= -1$. Several islands of stability immersed in the chaotic zone are shown. The arrows represent the positions of the islands of stability, while the numbers show the corresponding resonances along the line $p_{r}/m = 0$. Lower panel: rotation curve corresponding to the upper panel.
\label{fig6}}
\end{figure}

When the quadrupole moment is negative ($q/M^3 < 0$), the HT spacetime (\ref{HT-Metric}) corresponds to the prolate central object. However, for the prolate case, the structures of phase space do not change dramatically. There is a main island of stability, and surrounding it a chaotic sea of plunging orbits. In particular, a detail of the surface of section ($\theta=\pi/2, p_{\theta}>0$) for $q/M^3 = - 2$ is shown in Fig.~\ref{fig5}. A large island corresponding to the $2/7$-resonance is surrounded by several higher-ordered islands and chaotic regions. Most of the chaotic orbits presented in Fig.~\ref{fig5} are plunging orbits that remain attached to the stable orbits for a long period of time, before separating from the stable ones and plunging, due to the growing non-linearities. The chaotic orbits stick around the higher-ordered islands before they plunge. In particular, islands of stability are either embedded in a prominent chaotic layer or enveloped between KAM curves. The islands of higher multiplicity, e.g., the islands of stability $9/26$ and $6/17$ are embedded in the regular zone. 

The rotation curve corresponding to the PSs shown in the upper panel of Fig.~\ref{fig5} is presented in the lower panel. The prominent plateaus in the rotation curve belong to the $2/7$ (embedded in the chaotic zone) and $6/17$ (immersed in the regular region) resonant islands. However, there are several narrow plateaus surrounded by rapidly fluctuating intervals of rotation numbers. The rotation curve experiences an abrupt change at the hyperbolic point $r/M\approx4.948$,  after which it increases monotonically until reaching the $6/17$-resonance. Subsequently, beyond the $6/17$-resonance, the rotation curve continues to grow as a strictly monotonic function.      

If we change the value of the quadrupole moment parameter to $q/M^3= -1$, we do not see a dramatic change in the structure of the phase space. There exist chaotic sticky orbits near the outer boundary of the main island, and Birkhoff chains appear inside the main island of stability, see upper panel of Fig.~\ref{fig6}. However, the separation between the regular and chaotic regions is not so clear. Several thin islands of higher multiplicity, labeled with the corresponding resonances, immersed in the chaotic region can be observed along the line $p_{r}/m = 0$. The corresponding rotation curve indicates significant fluctuations when it crosses the initial conditions associated with chaotic orbits, see lower panel of Fig.~\ref{fig6}. The rotation curve takes the form of a plateau when crossing resonant islands of stability, and it changes abruptly when crossing the unstable periodic points of relatively small resonances. The island corresponding to $6/23$-resonance is embedded in the highly chaotic zone. Fig.~\ref{fig6} is a good example to note again that the islands of stability with lower multiplicity are more prominent than the higher multiplicity ones. Similar behaviour has also been observed for Zipoy-Voorhees metric \cite{Gerakopoulos:2012:PhRvD:}. Lower multiplicity islands of stability are very important since they are good candidates for detecting non-Kerr objects by the analysis of the gravitational waves coming from extreme mass ratio inspirals, even if the inspiraling smaller compact object might cross infinite resonances in a bumpy spacetime background during its inspiral \cite{Apostolatos-etal:2009:PhRvL:}.

\begin{figure}[h!]
\centering
\includegraphics[width=\hsize]{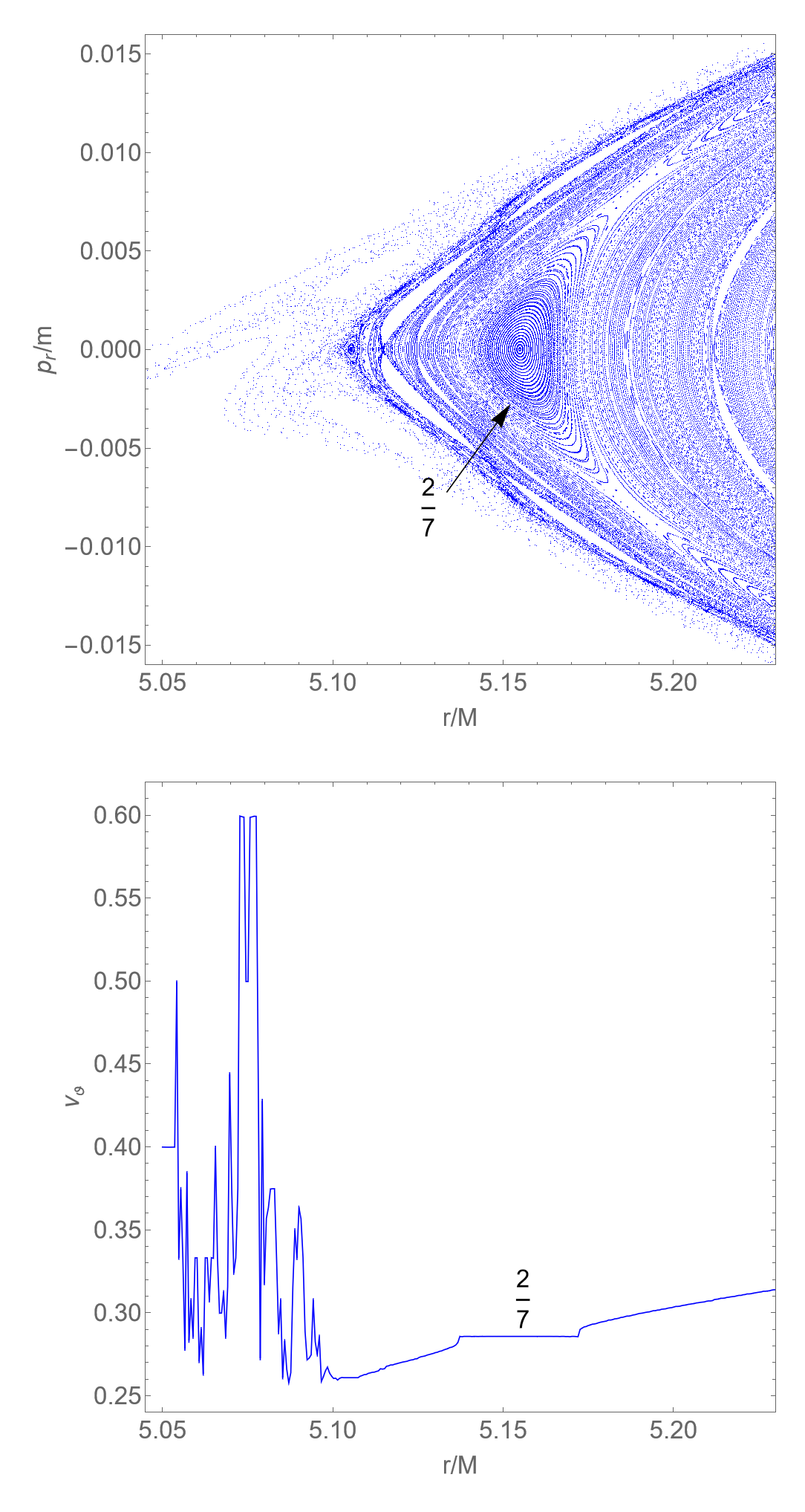}
\caption{Upper panel: a detail of the PS for parameters $\mathcal{L}/M=0.999$, $\mathcal{E}=0.95$, $J/M^2 =0.4$, and $q/M^3= -1.7$. An island of stability corresponding to resonance $2/7$ is shown. Lower panel: rotation curve corresponding to the upper panel. \label{fig8}
}
\end{figure}

HT metric quadrupole moment, responsible for deviation from regular dynamics, is stronger close to the neutron star's surface. Consequently, a higher density of PS structures is expected in the strong gravity region close to the neutron star's surface. Another example of the evolution of the structures in the PSs for $q/M^3<0$ is shown in Fig.~\ref{fig8}. The main island of stability is surrounded by a chaotic layer, where many high-multiplicity islands of stability are present. The boundary between this chaotic layer and the region occupied by regular orbits is densely populated by chaotic orbits. Some chaotic orbits within the chaotic sea surrounding the main island of stability will plunge to the neutron star's surface, while some chaotic orbits emerging between the islands of stability within the main island will remain non-plunging. These orbits that remain close to the boundary of the main island exhibit stickiness, see the upper panel of Fig.~\ref{fig8}. On the left side of the lower panel of Fig.~\ref{fig8}, the irregular variations of the rotation number confirm the chaoticity of the orbits that surround the main island of stability. The hyperbolic point at $r/M \approx 5.115$ separates the regular and chaotic regions. There are several thin plateaus surrounded by rapidly oscillating intervals of the rotation number. On the right of these irregular variations, the rotation curve seems to be strictly monotonic until we reach the $2/7$ resonant island. On the right of the $2/7$ resonant island, the rotation number seems to grow like a strictly monotonic function again. 

The chaotic orbits belonging to the chaotic thin layers, such as the ones described above, tend to stick near regular orbits and have significant effects on their frequency spectrum. As long as the orbits stick close to the regular orbits, they tend to acquire two main fundamental frequencies similar to those frequencies corresponding to the regular orbits. When the chaotic orbits move away from the regular orbits, they lose these two main fundamental frequencies, and their frequency spectrum becomes dominated by chaotic noise. However, these orbits can once again approach a regular trajectory, even the one they initially diverged from, stick around it for a specific period of time, and once more display two frequencies. If these regular orbits are associated with resonances, the ratio of their two fundamental frequencies remains a rational number for a certain period; otherwise, it becomes irrational. The appearance and disappearance of the two main frequencies characterize the existence of sticky chaotic orbits, thus indicating the existence of non-Kerr spacetime. However, detecting this phenomenon is challenging due to the interference of instrumental noise in gravitational wave signals. Therefore, the observational approach primarily emphasizes studying regular orbits corresponding to resonances \cite{Gera-etal:2010:PhRvD:}.

\begin{figure*}[!ht]
 \centering
\includegraphics[width=0.325\hsize]{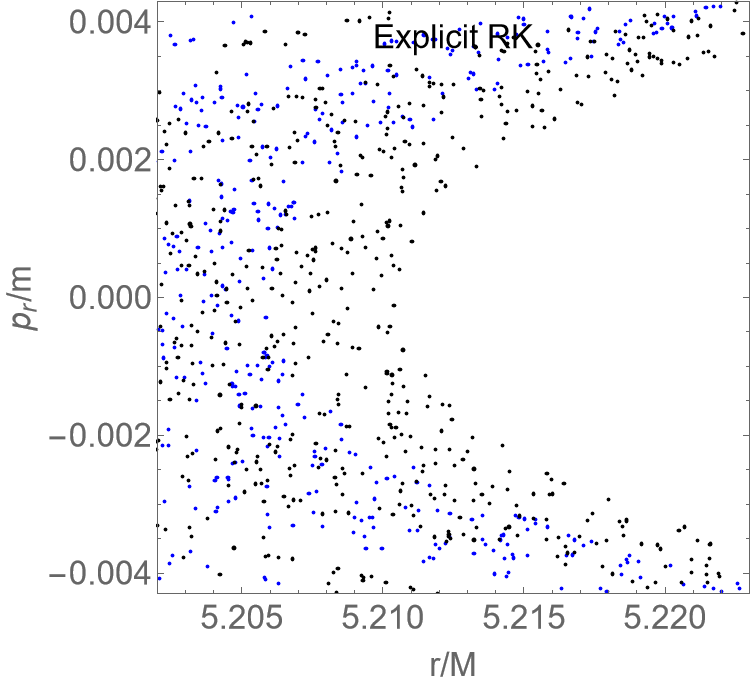}
\includegraphics[width=0.325\hsize]{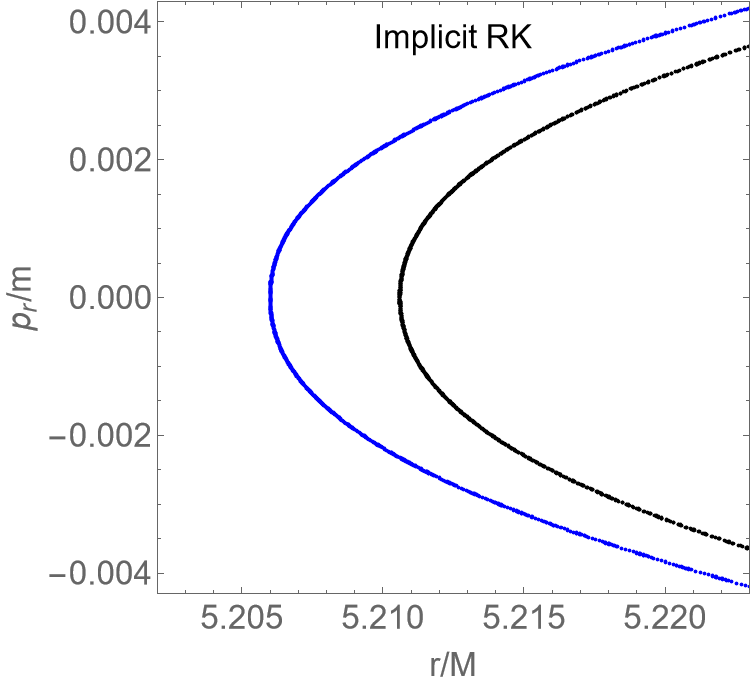}
\includegraphics[width=0.325\hsize]{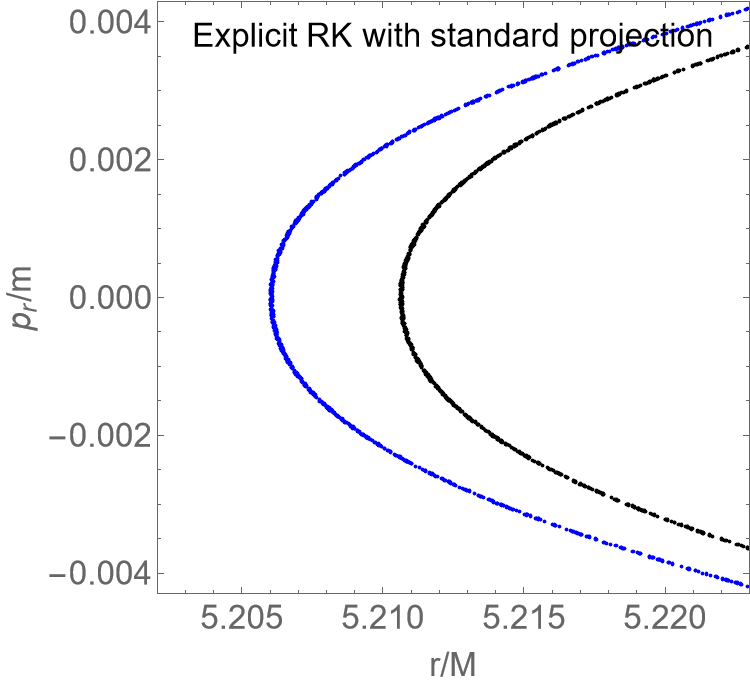}
\includegraphics[width=0.325\hsize]{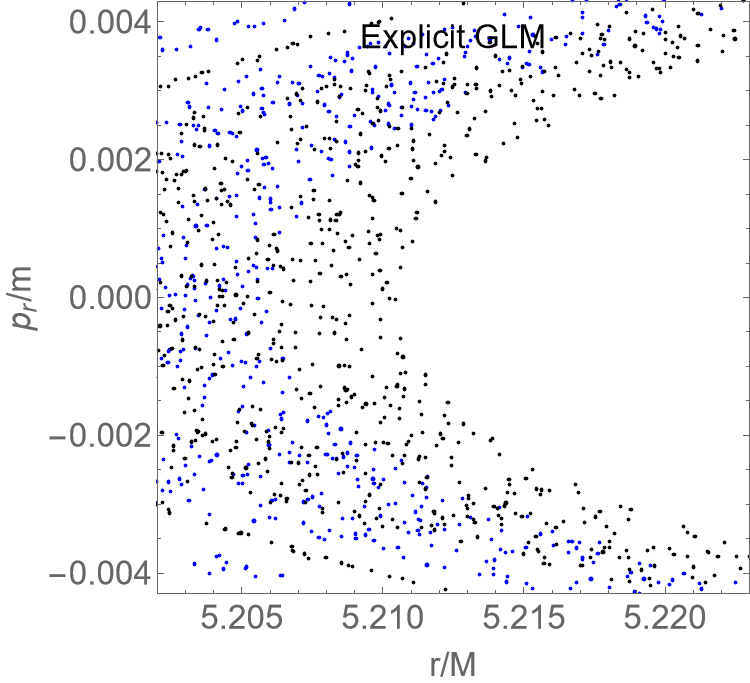}
\includegraphics[width=0.325\hsize]{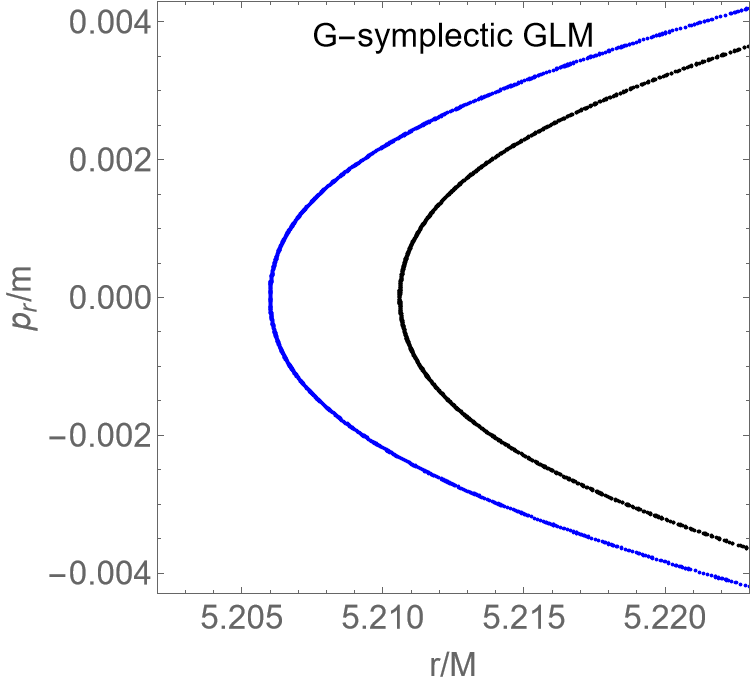}
\includegraphics[width=0.325\hsize]{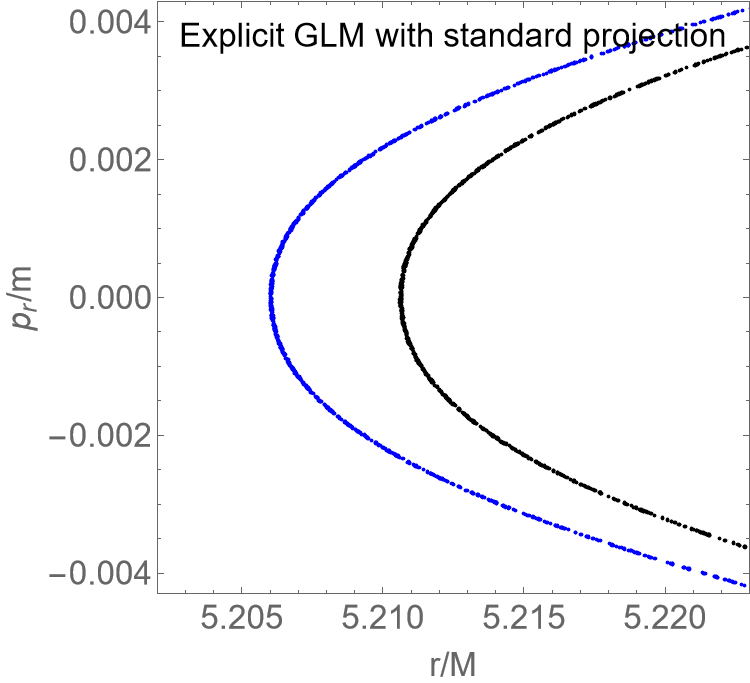}
\caption{PSs for two regular trajectories for $\theta = \pi/2$, $p_{r}/m = 0$, $p_{\theta}>0$, and parameters $\mathcal{L}/M=0.6795$, $\mathcal{E}=0.92579$, $J/M^2 = 0.4$, and $q/M^3=-1$, plotted using various numerical integrators of order four with integration time $\tau=5\times10^{5}$, and time step $\Delta\tau=0.1$. The text with each plot indicates the specific numerical integrator employed to plot the depicted trajectories. We demonstrate how the accuracy of the numerical integration significantly affects the appearance of the PS. 
\label{integrators}
}
\end{figure*}

\begin{figure*}[!ht]
\centering
\includegraphics[width=0.32\hsize]{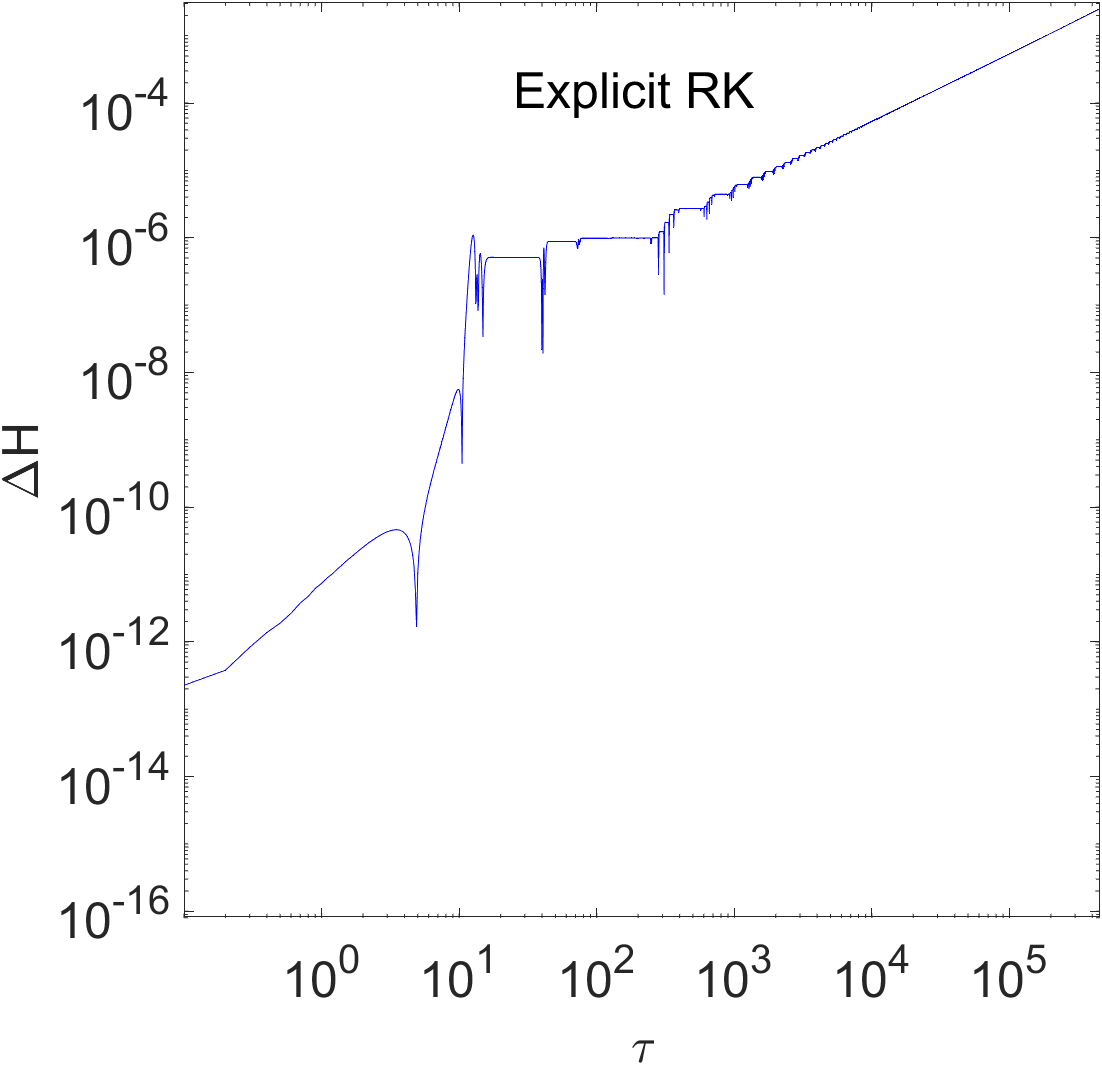}
\includegraphics[width=0.32\hsize]{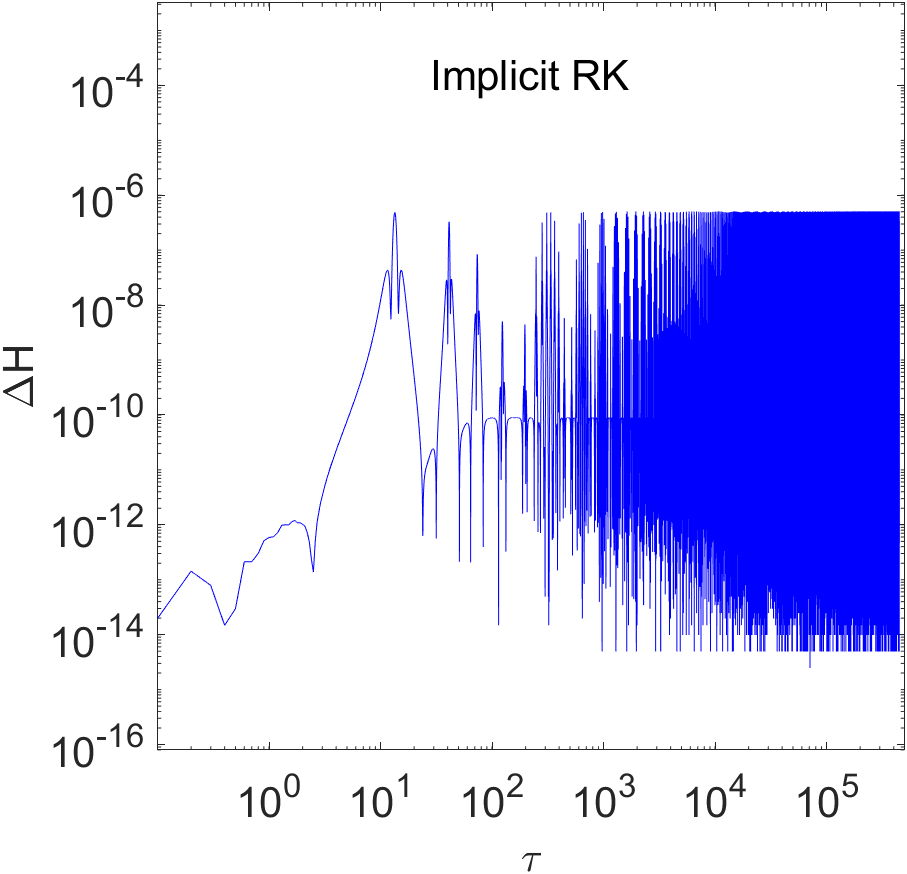}
\includegraphics[width=0.32\hsize]{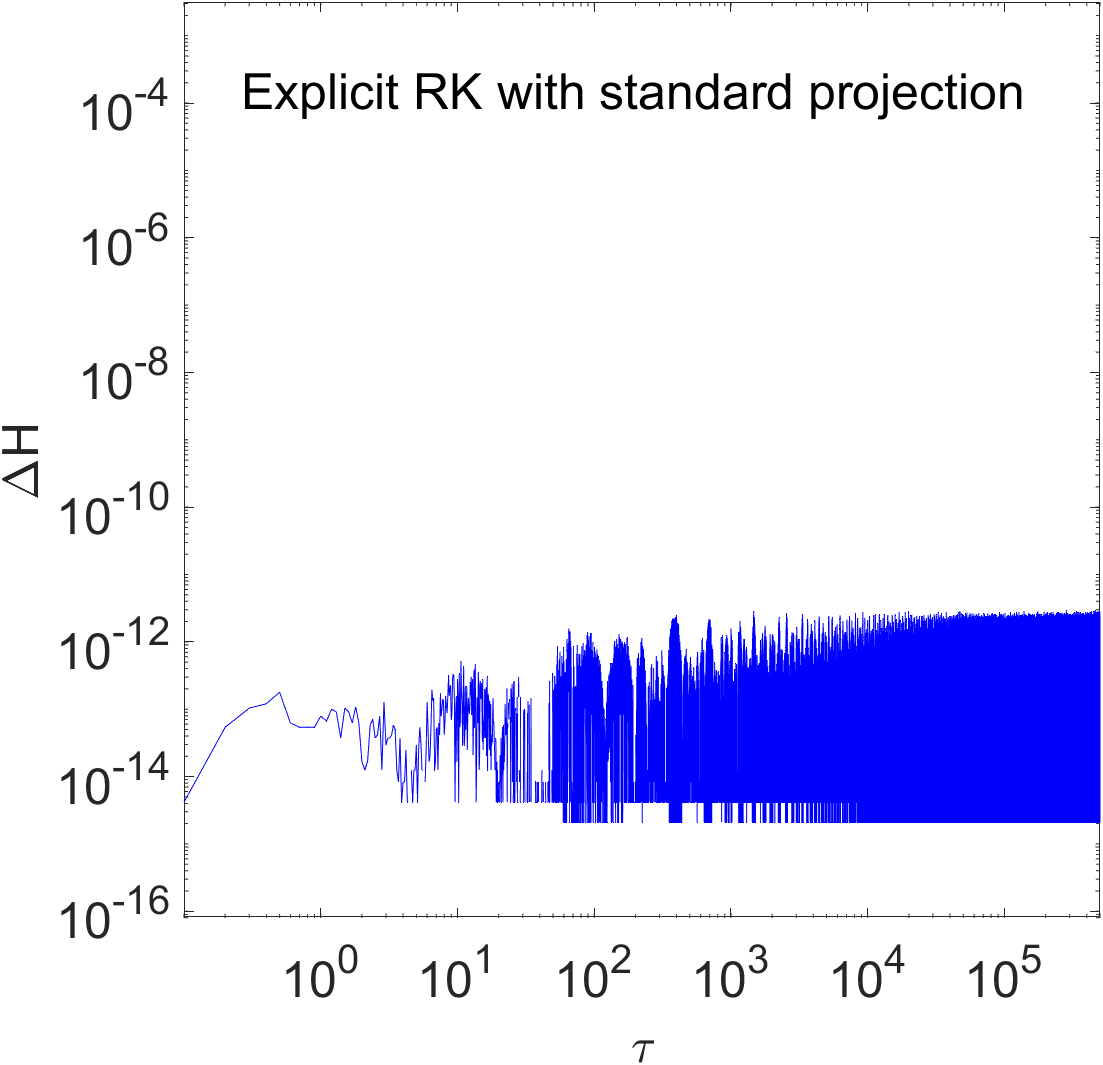}
\includegraphics[width=0.32\hsize]{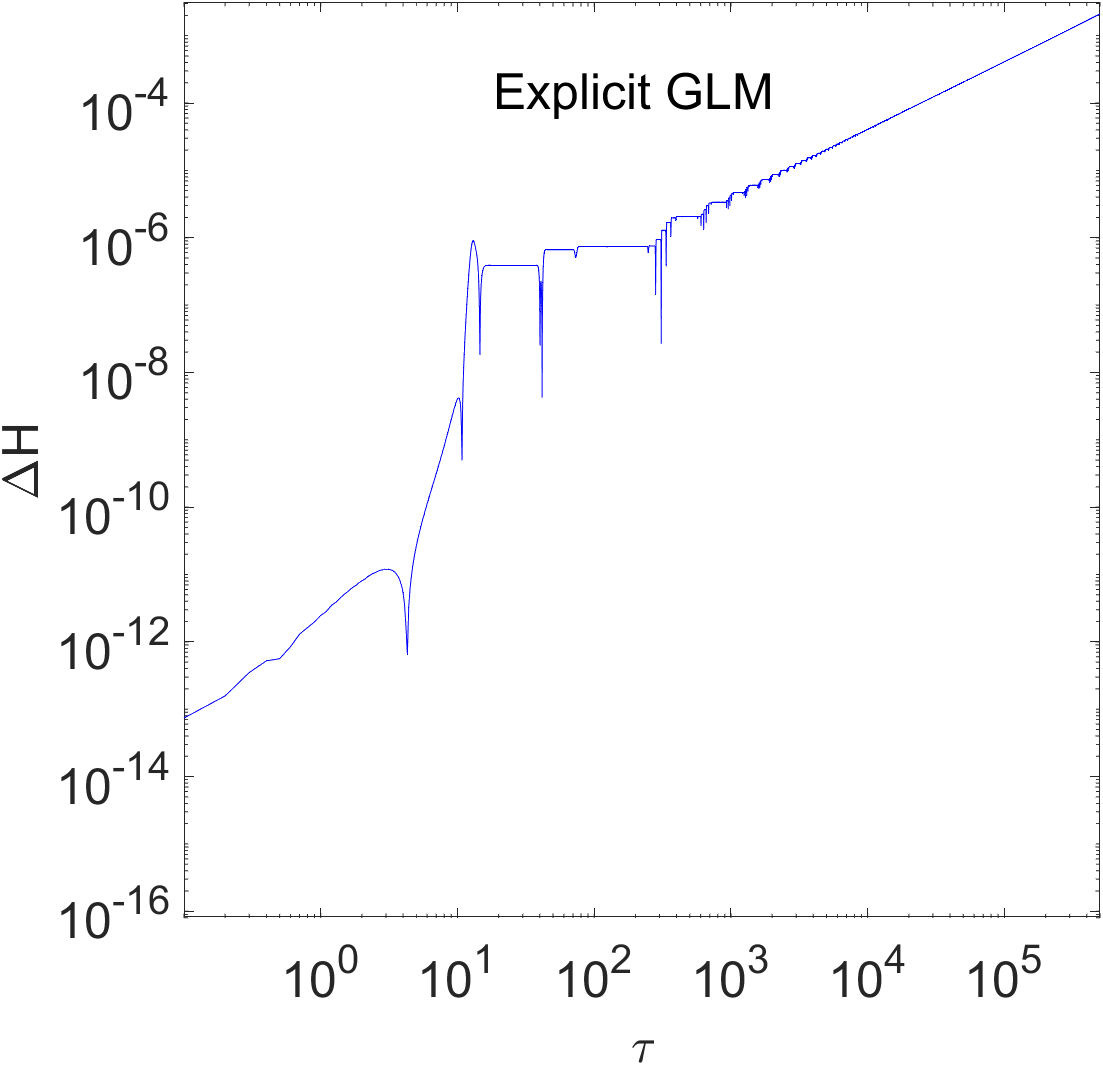}
\includegraphics[width=0.32\hsize]{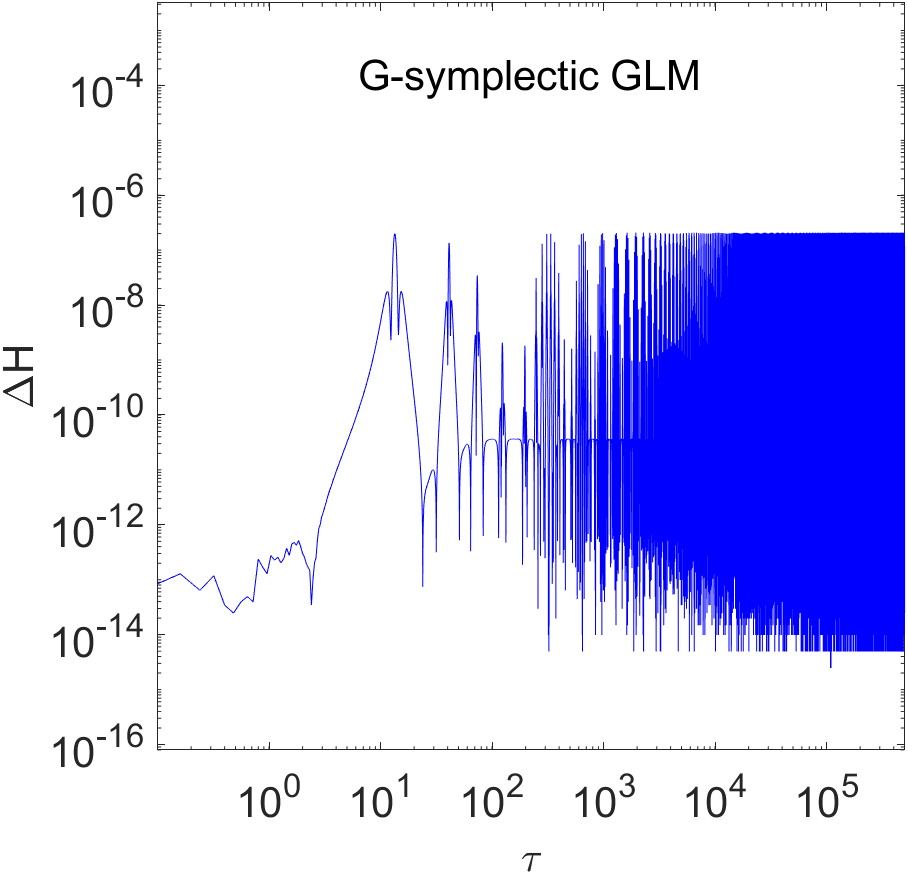}
\includegraphics[width=0.32\hsize]{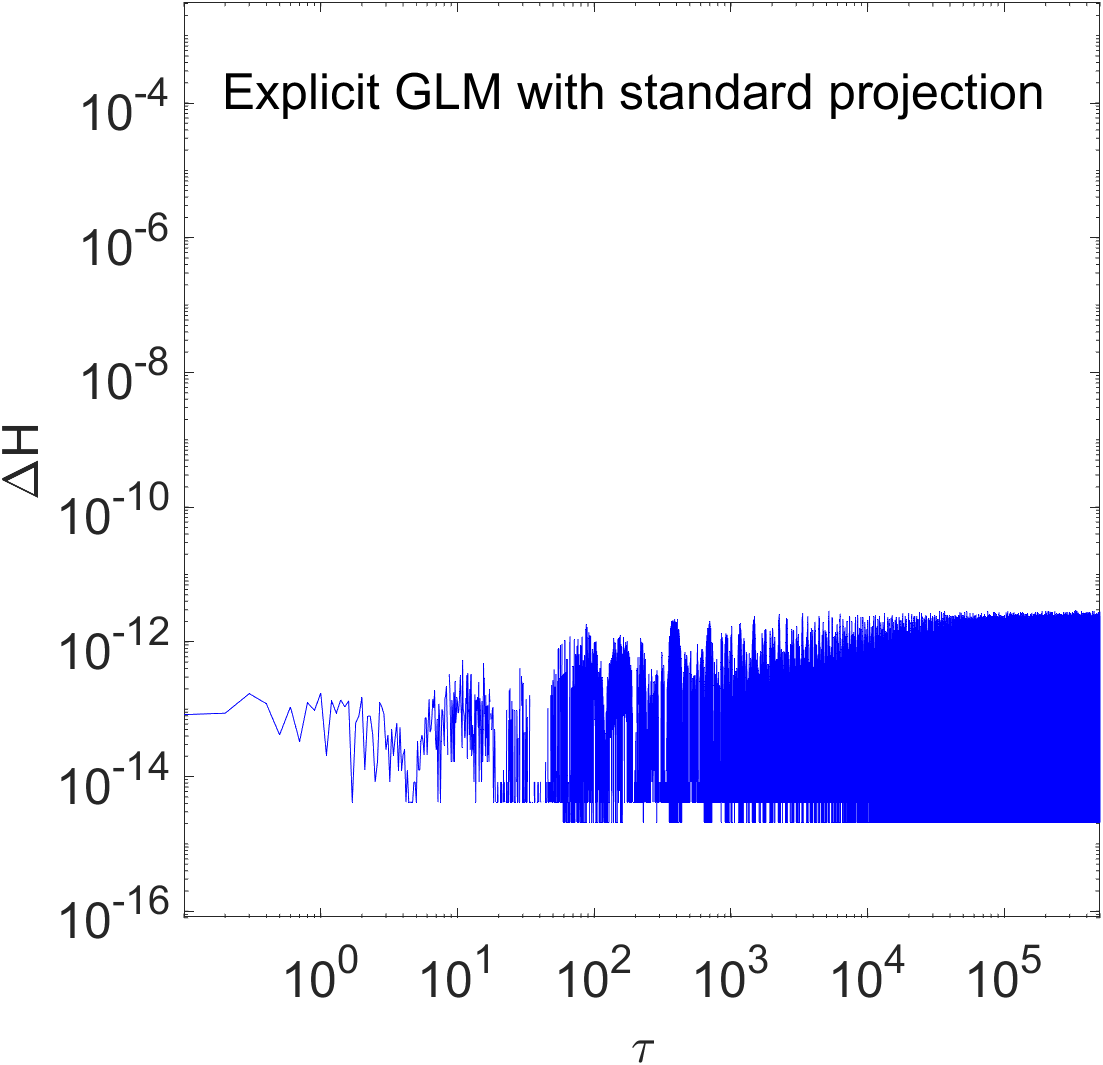}
\caption{Logarithmic plots showing the relative error in Hamiltonian for several numerical integrators employed to generate the regular trajectory (blue) presented in Fig.~\ref{integrators}. We demonstrate the growth of numerical errors in the Hamiltonian when employing different numerical integrators over extended time intervals.
\label{errors}
}
\end{figure*}

\begin{figure*}[!ht]
\centering
\includegraphics[width=0.325\hsize]{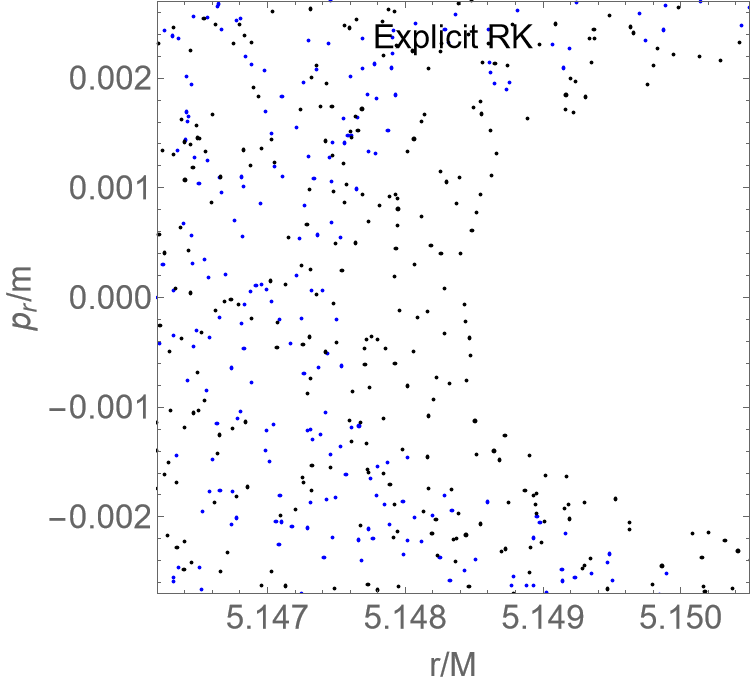}
\includegraphics[width=0.325\hsize]{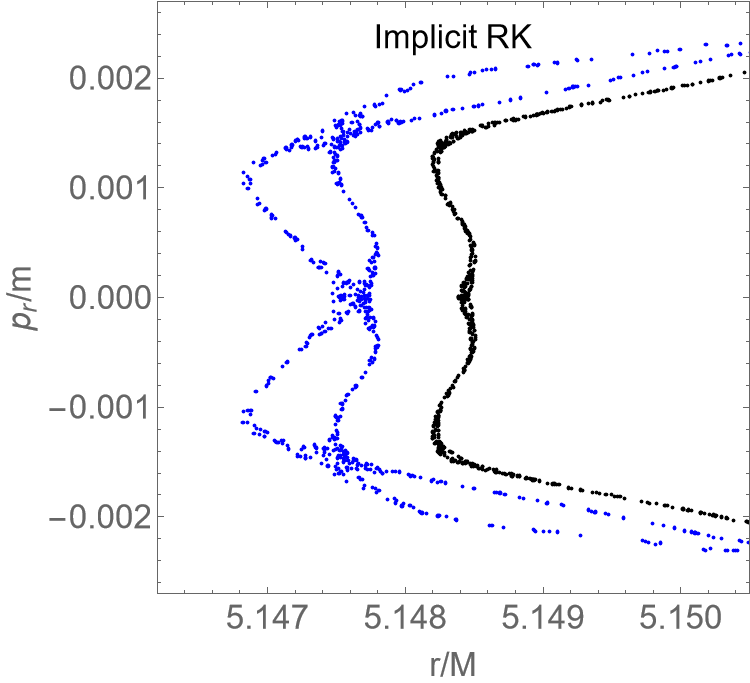}
\includegraphics[width=0.325\hsize]{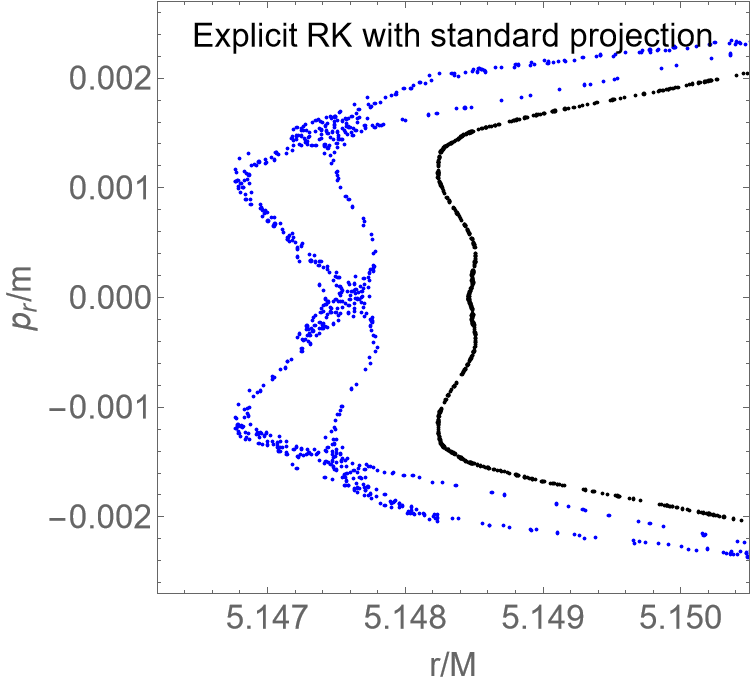}
\includegraphics[width=0.325\hsize]{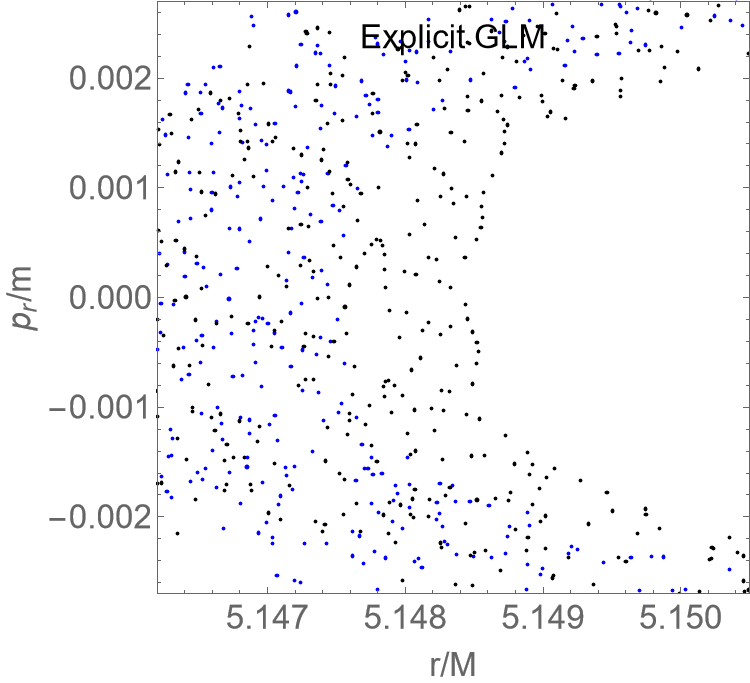}
\includegraphics[width=0.325\hsize]{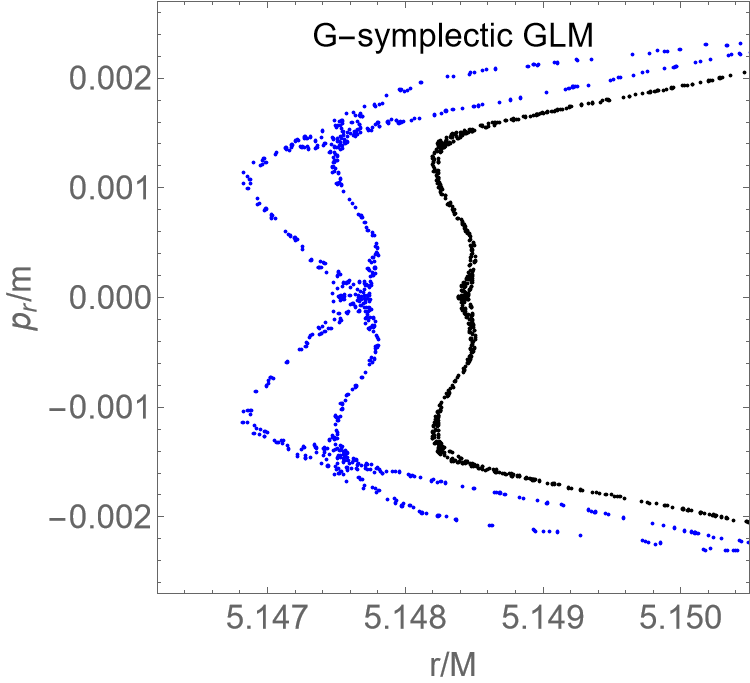}
\includegraphics[width=0.325\hsize]{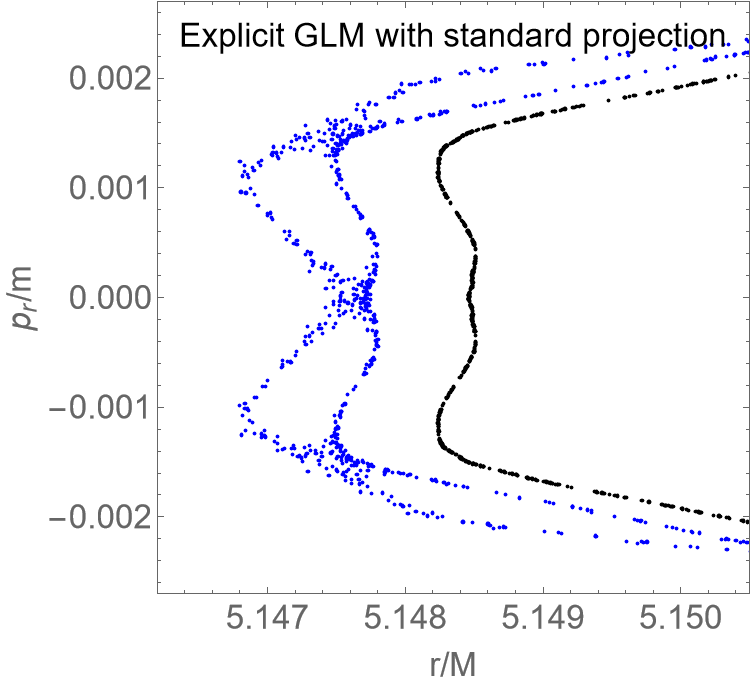}
\caption{PSs for two chaotic trajectories with parameters $\theta = \pi/2$, $p_{r}/m = 0$, $p_{\theta}>0$, $\mathcal{L}/M=0.6795$, $\mathcal{E}=0.92579$, $J/M^2 = 0.4$, and $q/M^3=-1$, plotted using various numerical integrators of order four with integration time $\tau=5\times10^{5}$, and time step $\Delta\tau=0.1$. The text with each plot indicates the specific numerical integrator employed to plot the depicted trajectories. We demonstrate how the accuracy of the numerical integration significantly affects the appearance of the PS. 
\label{integrators_2}}
 \end{figure*}

\begin{figure*}[!ht]
\centering
\includegraphics[width=0.32\hsize]{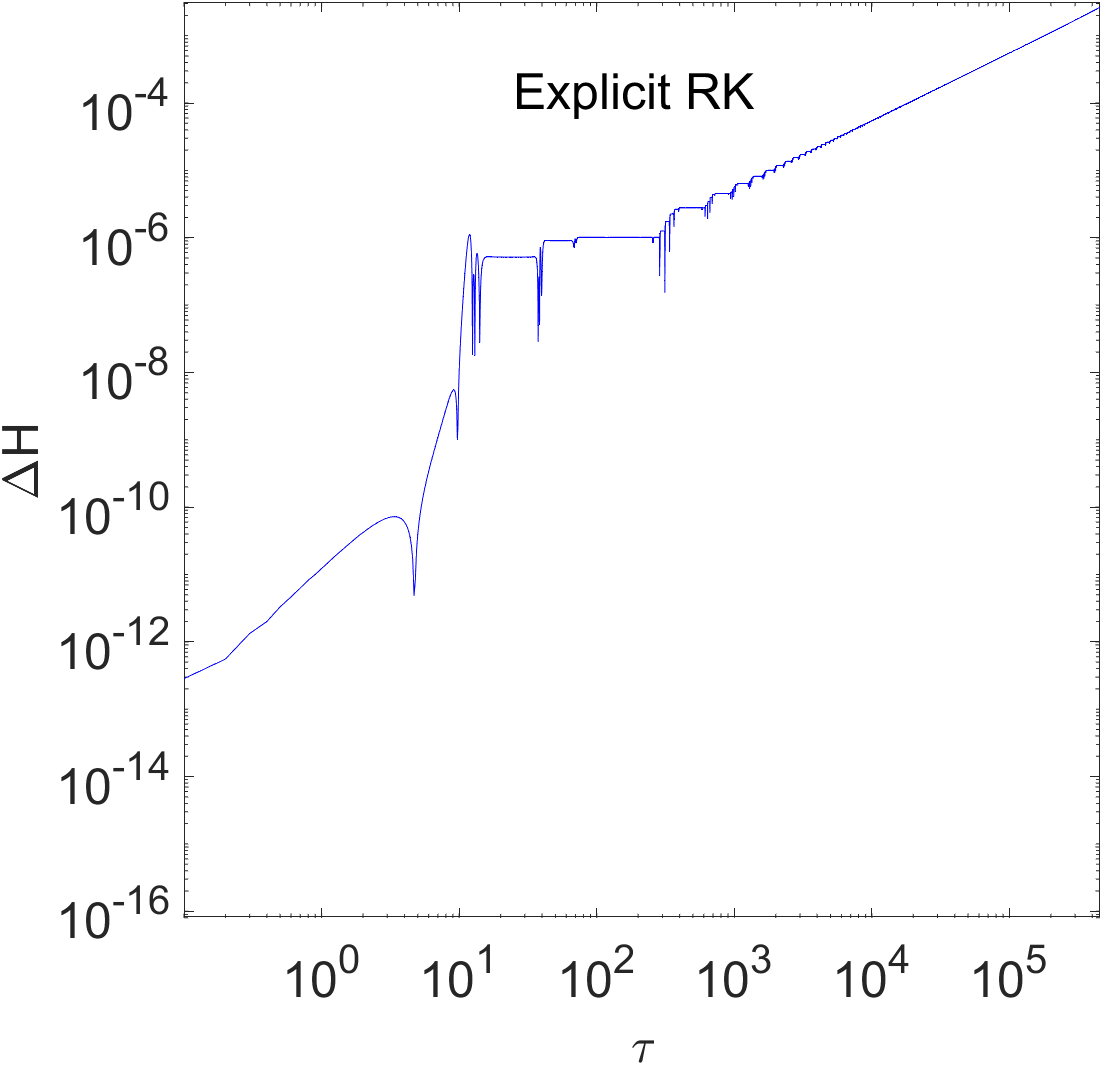}
\includegraphics[width=0.32\hsize]{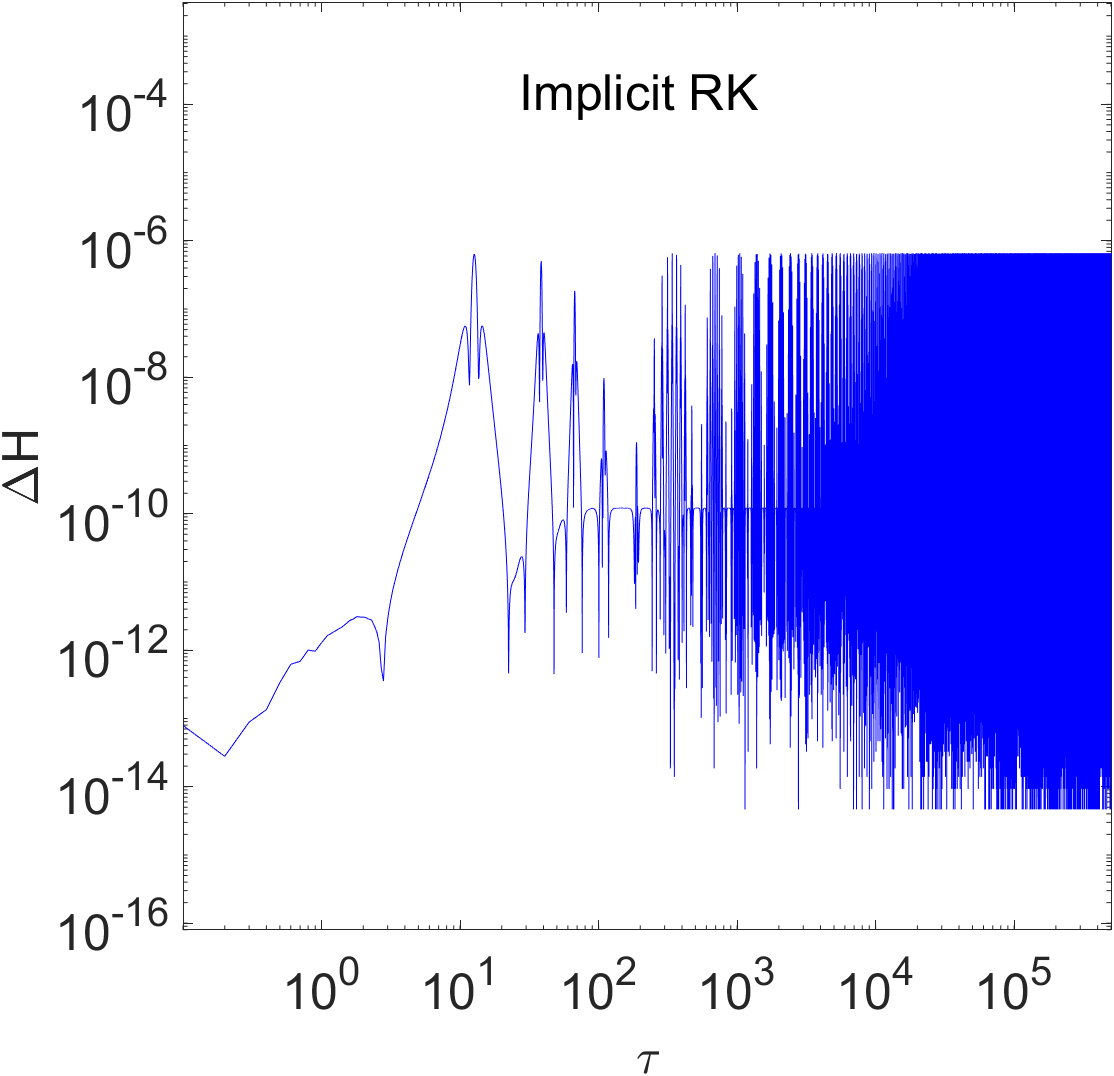}
\includegraphics[width=0.32\hsize]{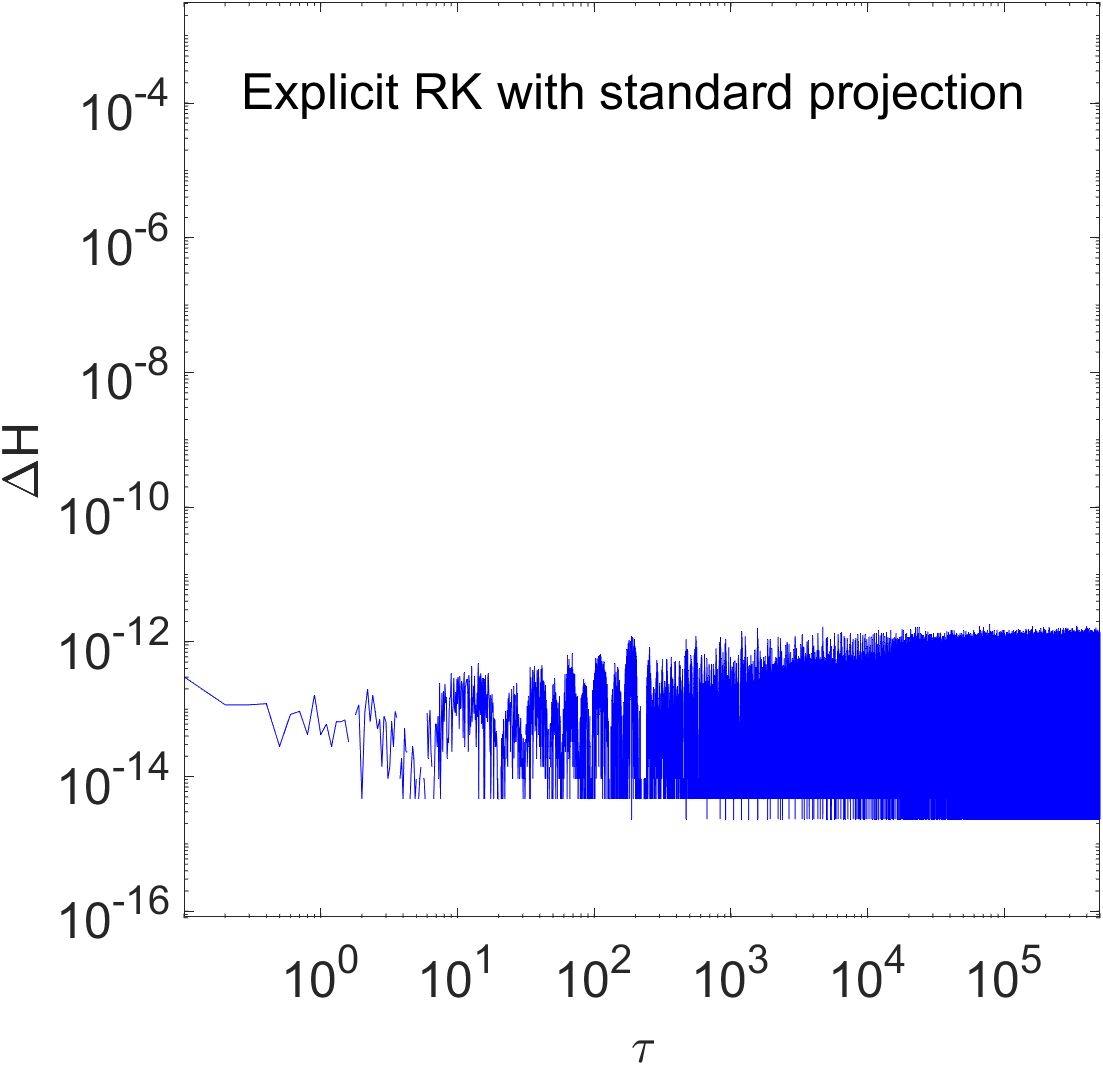}
\includegraphics[width=0.32\hsize]{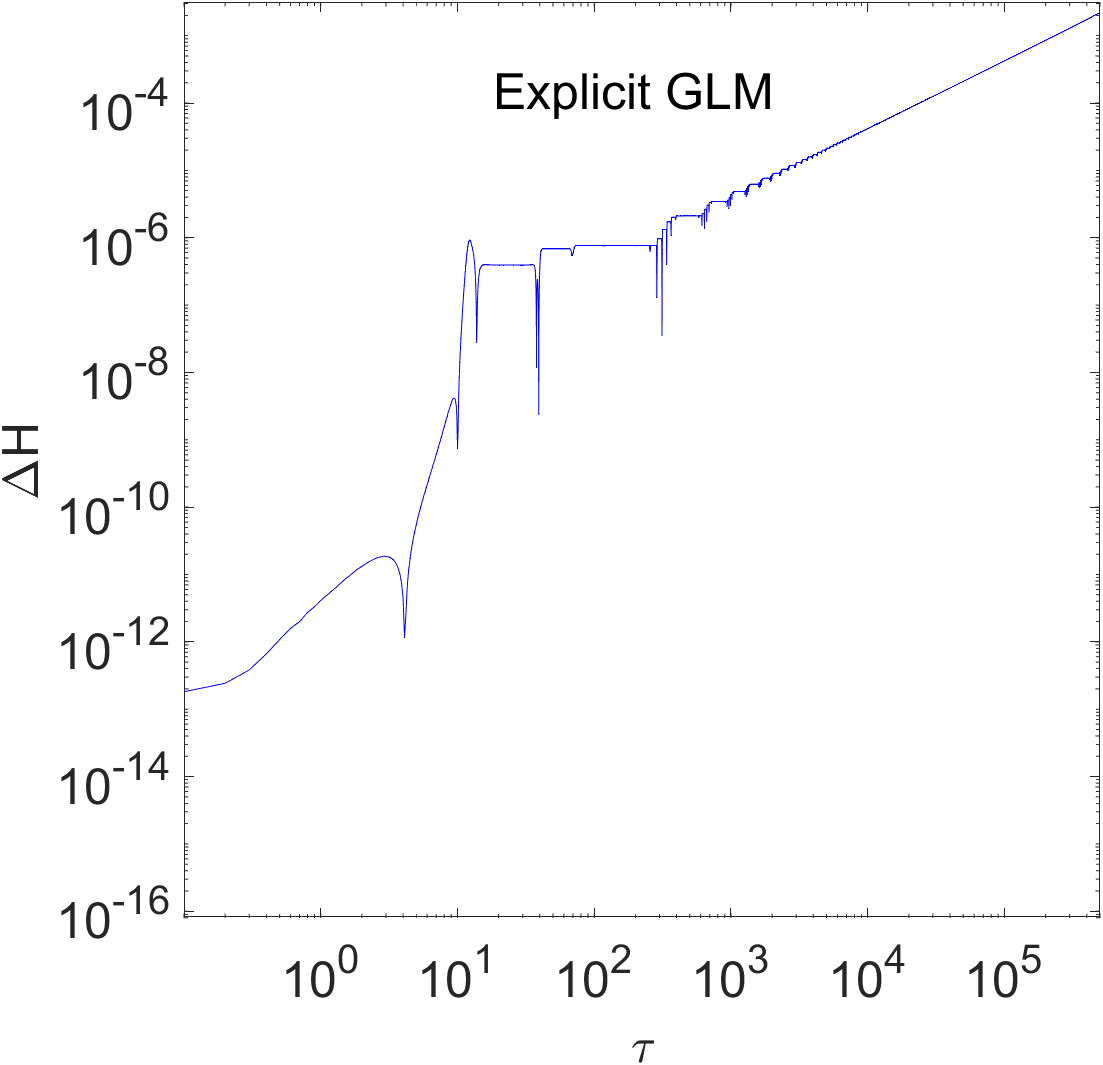}
\includegraphics[width=0.32\hsize]{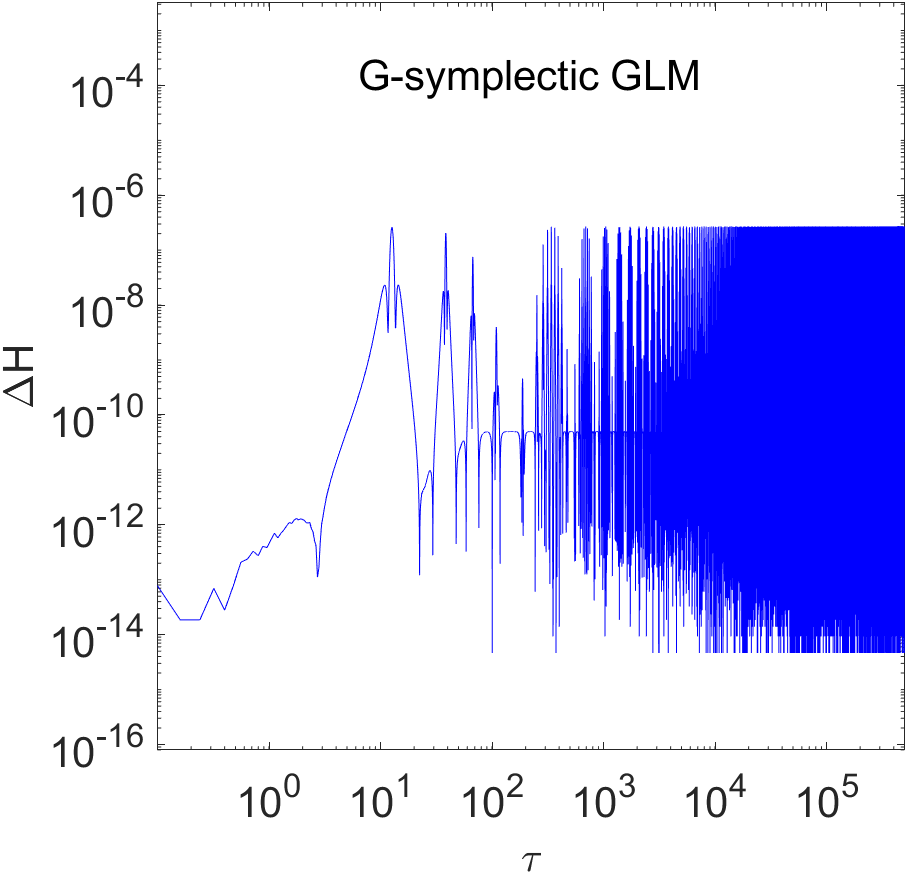}
\includegraphics[width=0.32\hsize]{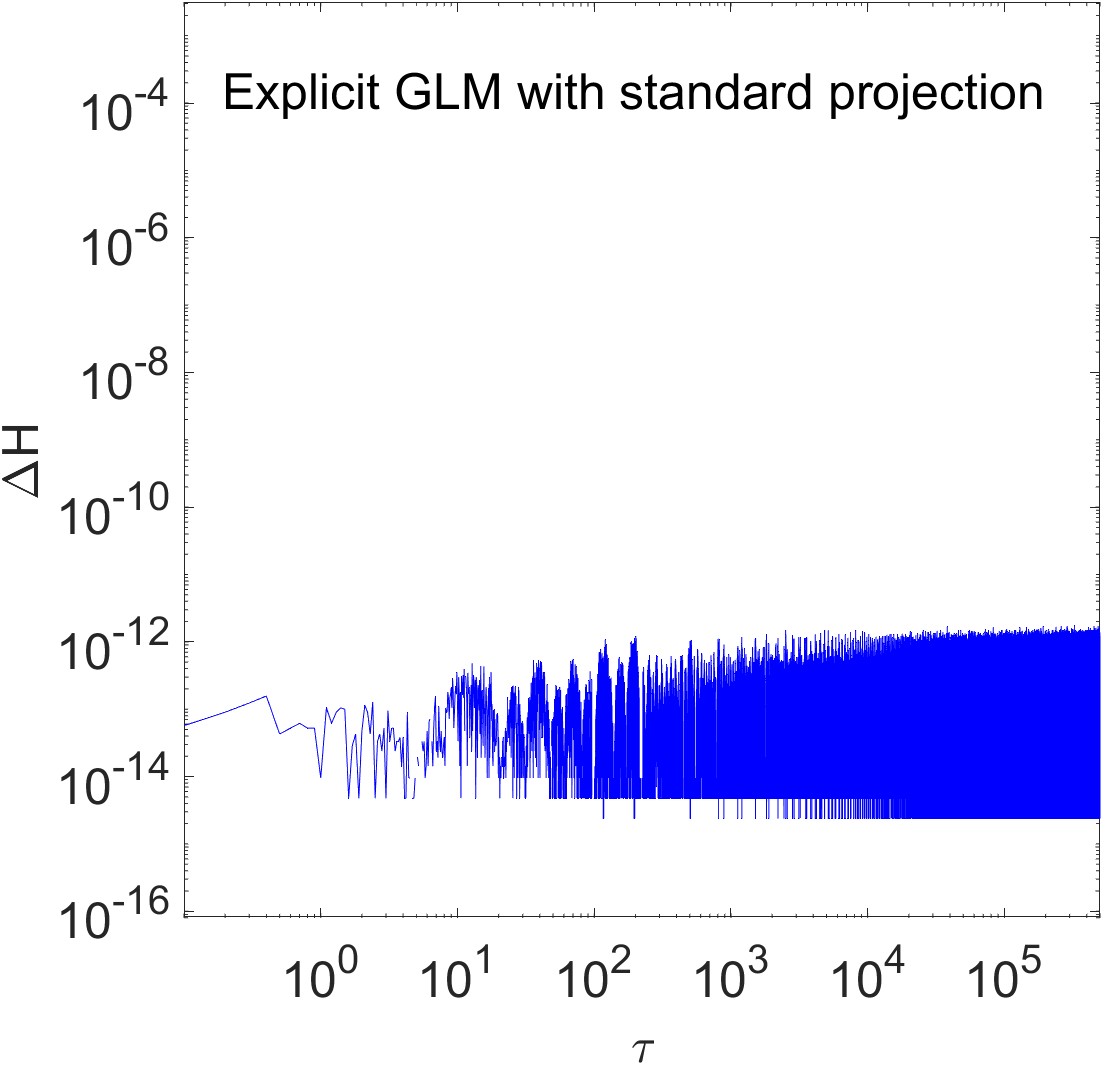}
\caption{Logarithmic plots showing the relative error in Hamiltonian for several numerical integrators employed to generate the chaotic trajectory (blue) presented in Fig.~\ref{integrators_2}. We demonstrate the growth of numerical errors in the Hamiltonian when employing different numerical integrators over extended time intervals.
\label{errors_2}}
\end{figure*}

\section{Numerical integrators and their performance}

Numerical methods approximate the exact solutions and hence introduce numerical errors. The error at each time step is usually known as discretization error or local truncation error, which accumulates over the course of the numerical integration and results in the global error \cite{Harier:2000:Book}. Numerical methods that produce small global errors have always been a preferred choice, but they do not always respect the qualitative features (consistency, stability, convergence) of the problem. The qualitative features of numerical methods, i.e., consistency, stability, and convergence, have always been the desired and often required goal of these numerical integrators. A numerical method is of no use if the numerical solution does not converge to the exact solution during the course of time. All these criteria focus on obtaining the quantitatively correct numerical solutions of the ODEs. However, there exist, classes of ODEs, where the qualitative behaviour of the solution is also important along with accuracy.

Geometric numerical integration is a specific class of numerical methods that take into account the geometric properties of the ODEs being solved. These methods are designed to preserve certain geometric properties of the exact solution, such as energy or momentum, which can be lost in traditional numerical methods. These methods are typically based on the idea of symplectic integration, which is a numerical technique that conserves the symplectic structure of Hamiltonian systems. Hamiltonian systems are a special class of ODEs that are used to describe the behaviour of many physical systems, such as planetary motion or the motion of particles around BHs and neutron stars. By preserving the symplectic structure of these systems, geometric numerical methods can provide highly accurate and stable solutions that are free from spurious oscillations or other numerical artifacts \cite{Sanz-Jesus:1992:,Sanz-Serna:2018:numerical:}. A more comprehensive discussion of the numerical integrators employed in this article is provided in the Appendix.

The performance of numerical integrators plays a crucial role in accurately solving mathematical problems and simulating dynamical systems. In this section, we will evaluate and analyse the performance of various numerical integrators to assess their efficiency, accuracy, and stability in simulating the motion of test particles around slowly rotating neutron stars.  

We present the simulations with several numerical integrators, namely, explicit RK, implicit Rk, explicit RK with standard projection, explicit GLM, G-symplectic GLM, and explicit GLM with standard projection, given by neutral test particle motion around a slowly rotating neutron star, shown in Fig.~\ref{integrators} for two regular trajectories. We use all integrators of order four and let the simulation run up to $\tau = 5\times10^5$ with a time step $\Delta\tau = 0.1$. The PSs exhibit variations in appearance with different numerical integrators. The implicit methods (both RK and G-symplectic GLM) are symplectic and preserve the regular structure of PSs. On the other hand, explicit methods (both RK and GLM), and explicit methods with standard projection which are non-symplectic and regular structures in PSs are corrupted, which could lead to the incorrect interpretation of trajectories as chaotic ones. Explicit methods with standard projection do not generally preserve the structure of PSs. However, by using a small time step, the structure of PSs can be preserved approximately.  

The accuracy of numerical integrators can be measured by the relative error in Hamiltonian (\ref{Hamiltonian}), given by 
\beq\label{H_error}
\Delta H(\tau) = \left| 1 - \frac{H(\tau)}{H(0) } \right|,
\eeq
and presented in Fig.~\ref{errors}, corresponding to the regular trajectories (blue) shown in Fig.~\ref{integrators}. We observe that the explicit methods with standard projection give much more precise results from the point of Hamiltonian error. The numerical errors in the Hamiltonian oscillate around zero, and the amplitude of the oscillations is bounded during the integration. Similar behaviour can be observed for implicit methods as well. Although we observe a minor variation in the Hamiltonian error plots, the Hamiltonian error demonstrates bounded behaviour over extended time intervals, which is a typical behaviour of the symplectic numerical integrators. However, for the case of explicit integrators, the Hamiltonian is not preserved and errors in Hamiltonian increase monotonically with the integration time $\tau$, which will lead to the gradual corruption of structures in PSs.


\begin{table}[!ht]
\begin{center}
\begin{tabular}{ l | c c c l }
\hline
Integrators  & $N_{\rm FE}$  & $ \norm{\Delta H}_{2}$  & symplectic \\
\hline
Explicit RK   & $5\times10^7$ & $0.018$ & No \\
Implicit RK & $2.6\times 10^8$ & $7.4\times10^{-7}$ & Yes \\
Explicit RK with\\ standard projection  & $7.5 \times 10^7$ & $1.7\times10^{-11}$ & No \\
Explicit GLM & $5\times10^7$ & $0.014$ & No \\
G-symplectic GLM   & $1.3\times 10^8$ & $7.7 \times 10^{-8}$ & Yes \\
Explicit GLM with\\ standard projection & $7.5 \times 10^7$ & $1.6\times 10^{-11}$ & No \\
\hline
\end{tabular}
\caption{Information on the performance of the numerical schemes used for the simulation of the regular trajectory (blue) around the slowly rotating neutron star, presented in Fig \ref{integrators}. All numerical integrators of order four, simulations time $\tau = 5 \times 10^{5}$ and time step $\Delta \tau = 0.1$ has been used. The number of function evaluations $N_{\rm FE}$, and $l_{2}$-norm of the Hamiltonian error, are reported.
\label{tab1}
}
\end{center}
\end{table}


It is interesting to note that explicit RK (and GLMs) with standard projection maintain the preservation of the Hamiltonian but retain the structure of the PSs only partially. We have employed these explicit methods with a standard projection technique, specifically designed to ensure the preservation of the Hamiltonian. These methods only preserve the total energy of the system but do not preserve the symplectic structure or other invariants of the system. If we use these explicit integrators with symplectic projection, then the resulting integrator would have symplectic behaviour, successfully preserving both the Hamiltonian and the PSs \cite{kraus:2017:arXiv}.   

The comparison of the performance of several non-symplectic and symplectic numerical integrators is also shown in Tab.~\ref{tab1}. We use all integrators of order four, integration time $\tau=5\times10^{5}$, time step $\Delta\tau=0.1$, and report the number of function evaluations $N_{\rm FE}$ and $l_{2}$-norms of the Hamiltonian error. Implicit methods require a higher number of function evaluations $N_{\rm FE}$ compared to explicit methods, as presented in Tab.~\ref{tab1}. Specifically, the number of function evaluations $N_{\rm FE}$ for implicit methods is twenty times more than that of explicit methods. Thus, the implicit methods are computationally expensive, whereas the explicit methods are cost-effective. The $l_{2}$-norms of the Hamiltonian error for implicit methods, as well as explicit methods with standard projection, are quite small, thus these methods preserve the Hamiltonian, whereas explicit methods do not preserve the underlying structure including the Hamiltonian. 

The simulation for chaotic trajectories employed with several integrators is depicted in Fig.~\ref{integrators_2} and the corresponding relative errors in Hamiltonian are presented in Fig.~ \ref{errors_2}. Since the dynamical system we are considering in our study exhibits weak chaos, we do not observe significant differences in the numerical errors of the Hamiltonian corresponding to the regular and chaotic trajectories. The numerical errors for both regular and chaotic trajectories in a dynamical system consisting of a Kerr BH immersed in the uniform magnetic field have been examined in \cite{Kopack:2014:ragt.conf:}, where the strong chaotic behaviour is observed, and different numerical errors corresponding to regular and chaotic trajectories are reported.

\section{Discussion and Conclusions}\label{kecy}

We explore the chaotic dynamics of test particles in HT spacetime which characterizes the geometry around slowly rotating and deformed objects within strong gravitational fields. This spacetime provides a framework for studying real astrophysical compact objects ranging from celestial bodies like planets to neutron stars. It is characterized by three multipole moments, namely the total mass, the spin angular momentum, and the quadrupole moment. The quadrupole moment describes the deviation from the corresponding Kerr object.

We have shown by several numerical examples that the particle dynamics in HT spacetime for both prolate and oblate deformations exhibit characteristics of a non-integrable system. This observation has also been found for photon orbits \cite{Kos-Pap:2022:CQGra:} and massive particle orbits \cite{Kyriakos-Kostas:2023:arXiv230518522D:} in HT spacetime with prolate deformations. In our work, the PSs and rotation numbers have been employed to detect the indications of HT non-integrability, which are the ideal tools to determine the non-linear evolution of the system. Furthermore, the rotation curve serves as a reliable indicator of chaos as well. Therefore, even without employing a surface of section or any other method to identify chaos, the rotation curves themselves are sufficient to demonstrate the existence of chaos. This is interesting from an observational perspective as well because the rotation number represents the ratio between two fundamental frequencies of a non-plunging orbit. Consequently, the rotation number acts as a suitable tool for detecting chaotic phenomena in gravitational wave signals. 

Strong non-linear character of equations of motion in a region close to the neutron star's surface leads to the emergence of many structures in phase space such as chains of islands, chaotic points, hyperbolic points, and higher-ordered islands. These structures have been observed clearly at the tip of the main island of stability. The islands of stability with lower multiplicity are more prominent than the higher multiplicity ones. Similar behaviour has also been observed for Zipoy-Voorhees metric \cite{Gerakopoulos:2012:PhRvD:}. Lower multiplicity islands of stability are very important since they are good candidates for detecting non-Kerr objects by the analysis of the gravitational waves \cite{Apostolatos-etal:2009:PhRvL:}.     

The stickiness phenomenon plays a crucial role in describing the behaviour of chaotic geodesics, as they remain attached to stable geodesics for a long period of time before eventually diverging. We have found that most of the chaotic orbits exist near the outer boundary of the main island of stability, and Birkhoff chains appear inside the main island of stability. Some orbits remain attached to higher-ordered islands while others follow a trajectory leading them toward the neutron star's surface, due to the growing non-linearities. Similar results can also be found for non-Kerr spacetime \cite{Adrian-etal:2022:PhRvD:}.

We have found that the most prominent islands, characterized by their significant width, correspond to $1/3$-resonances. The widths of resonant islands are proportional to the spacetime parameters, however, the width of the resonances increases with the particle's energy $\mathcal{E}$ and decreases with the orbital angular momentum $\mathcal{L}/M$ \cite{Kyriakos-Kostas:2023:arXiv230518522D:}. For Manko-Novikov spacetime, the most prominent islands are of the smallest multiplicities, namely the $2/3$-resonance \cite{Gera-etal:2010:PhRvD:}.

We have investigated the structures of PSs in both prolate and oblate deformations. In our limited study, we do not observe significant differences in these structures between the two cases. Chaos is present in both scenarios, regardless of whether the HT spacetime corresponds to a prolate or oblate central compact object. Specifically, all orbits within the chaotic sea that surrounds the main island are sticky chaotic orbits. There are several thin islands of stability, hyperbolic points, and higher-ordered islands immersed in the chaotic region. The islands of stability with lower multiplicity are more prominent than the higher multiplicity ones. 

We presented a generalized framework for the numerical integration of test particle motion in the context of general relativity. We performed a detailed comparison between several fourth-order numerical integrators, namely the standard explicit RK method, explicit GLM, implicit Gauss RK method, implicit G-symplectic GLM, explicit RK method with standard projection technique, explicit GLM with standard projection technique, which are suitable for simulations of test particles.

We have compared the accuracy and performance of numerical schemes using numerical errors in the conserved Hamiltonian system consisting of HT spacetime. The precision of the numerical integration plays a crucial role in the appearance of structures in the PS which could be used for chaos detection. The implicit RK and G-symplectic GLM preserve the structural integrity of the PSs due to their symplectic nature. On the other hand, explicit integrators (RK and GLM) and explicit methods with standard projection, being non-symplectic, result in an inaccurate interpretation. In the case of implicit schemes and explicit integrators with standard projection, the numerical errors in the Hamiltonian oscillate around zero and the amplitude of these oscillations remains bounded. The errors exhibit a bounded pattern over extended time intervals, which is a distinctive characteristic observed in symplectic numerical integrators. In contrast, the numerical errors for explicit integrators (RK and GLM) increase steadily as the integration time $\tau$ is extended. Consequently, the explicit methods fail to maintain the integrity of the Hamiltonian structure.

We propose that our findings are not limited to a specific problem of test particle dynamics in HT spacetime, and can be extended to a wide range of systems within the same class. Our work can serve as a short survey from which one can choose a numerical integrator with the desired qualities. In particular, implicit methods (RK and GLM) maintain the preservation of both the Hamiltonian and PSs, whereas explicit methods do not preserve either. In contrast, explicit methods with standard projection ensure the preservation of the Hamiltonian while failing to maintain the integrity of the PSs, however, by using a small time step, we can approximately preserve the structure of PSs. 

Within the context of the presented numerical integrators, there exist several intriguing possibilities for future exploration, which include variational integrators, discrete gradient methods, and explicit schemes with alternative projection techniques. These extensions hold promise and could offer interesting directions for further investigation\footnote{Our codes are publicly available and can be used for further development and application in other non-linear dynamical systems \url{https://github.com/Scheherazaade/Chaotic_dynamics}}.

\section*{Acknowledgments}

This work is supported by the Research Centre for Theoretical Physics and Astrophysics, Institute of Physics, Silesian University in Opava, and Czech Science Foundation Grant No.~\mbox{23-07043S}.



\def\prc{Phys. Rev. C}
\def\pre{Phys. Rev. E}
\def\prd{Phys. Rev. D}
\def\jcap{Journal of Cosmology and Astroparticle Physics}
\def\apss{Astrophysics and Space Science}
\def\mnras{Monthly Notices of the Royal Astronomical Society}
\def\apj{The Astrophysical Journal}
\def\aap{Astronomy and Astrophysics}
\def\actaa{Acta Astronomica}
\def\pasj{Publications of the Astronomical Society of Japan}
\def\apjl{Astrophysical Journal Letters}
\def\pasa{Publications Astronomical Society of Australia}
\def\nat{Nature}
\def\physrep{Physics Reports}
\def\araa{Annual Review of Astronomy and Astrophysics}
\def\apjs{The Astrophysical Journal Supplement}
\def\aapr{The Astronomy and Astrophysics Review}
\def\procspie{Proceedings of the SPIE}

\bibliographystyle{spphys}
\bibliographystyle{unsrt}
\bibliography{reference}

\appendix 

\section{Numerical Integrators}

Numerical integration is a widely used technique for approximating the solution of ODEs when an exact solution is either impossible or very difficult to obtain analytically. Numerical methods can be categorized as one-step methods, multi-step methods, and GLMs. One-step methods approximate the solution of a differential equation using only one previous value. This means that the solution $y(x)$ at a particular point $x_{n}$ depends only on the previous value $y(x_{n-1})$ of the solution. However, multi-step methods require several previous values ($y(x_{n-1}), y(x_{n-2}), \cdots$) of the solution $y(x)$ to approximate the value at a particular point $x_{n}$, thus the solution at a particular point depends on several previous values of the solution. Similarly, the GLM also requires several input values to start the procedure. These methods are implemented in a recursive way, and a starting method is generally employed in order to start the procedure. Usually, one-step methods such as RK methods are used as starting methods. When the data is available to start the procedure, a multi-step method is then employed. Multi-step methods tend to be more accurate and more stable than one-step methods, but they are also more complex to implement. One-step methods are easy to implement, but multi-step methods can be more computationally expensive than one-step methods \cite{Eirola-etal:1992:conservation:, Butcher:2006:GLM:,glm_underlying_onestep_method}.

Numerical integrators can be classified into explicit and implicit ones. Explicit methods calculate the solution at the next time step solely based on the solution at the current time step (without using any information from the next time step), while implicit methods use information from both the current and next time steps to calculate the solution. Both schemes have their own advantages and limitations. Explicit methods are generally easier to implement and computationally efficient, but they may become unstable if the time step is too large or if the differential equation is stiff or highly non-linear. These methods are commonly used for solving simple or moderately complex problems where accuracy is not critical. However, implicit methods are generally more accurate and stable than explicit methods for stiff or highly non-linear problems, but they require more computational resources and are more difficult to implement. Implicit methods are commonly used for solving complex problems that require high accuracy and stability \cite{Sanz-Jesus:1992:,Sanz-Serna:2018:numerical:}.

Explicit schemes have a major limitation when it comes to Hamiltonian systems. They do not preserve the Hamiltonian exactly, and the errors in energy conservation can accumulate over time. On the other hand, implicit schemes are more accurate and capable of preserving the Hamiltonian exactly. In the following, we discuss various numerical integrators based on their ability to preserve the qualitative features of numerical integrators.

\subsection{Implicit structure-preserving schemes}

Implicit structure-preserving schemes are numerical integration methods that are designed to accurately simulate the behaviour of physical systems by preserving their underlying structure. These schemes are particularly useful for systems with complex or non-linear dynamics and can be applied to both Hamiltonian and non-Hamiltonian systems. By preserving the underlying structure, implicit schemes ensure that important physical properties like energy conservation, symplecticity, and momentum conservation are maintained throughout the simulation. One of the key advantages of these schemes is their ability to accurately simulate long-term behaviour, particularly in systems that exhibit chaotic or oscillatory behaviour. These methods are well-suited for systems with stiff differential equations, which can be difficult to solve using explicit methods. Implicit methods can be applied to stiff systems because they do not require small-time steps to ensure numerical stability, unlike explicit methods \cite{Harier:2000:Book,habib:2010:thesis}.

Symplectic algorithms are numerical integrators of Hamiltonian systems that preserve the symplectic structure in phase space. In long-term integration, these algorithms tend to perform better than their non-symplectic counterparts. An additional benefit of such methods is their ability to preserve the underlying quadratic invariants effectively. However, most of the numerical methods in practice are not symplectic. Multi-step methods require more than one initial condition to start with, thus they can not define a map on phase space and hence cannot be symplectic in general. The one-step methods for the numerical integration of Hamiltonian systems are said to be canonical or symplectic if when applied to any Hamiltonian systems with any step length they give rise to a symplectic transformation in phase space. It is well known that implicit schemes can only be considered as truly symplectic, whereas explicit schemes do not preserve the flow of the Hamiltonian systems \cite{Sanz-Jesus:1992:}.

\subsubsection{Symplectic RK methods}

RK methods are a family of one-step numerical algorithms commonly used to approximate the solutions of initial value problems (IVPs)
\begin{equation}
y'(x)=f(y(x)), \hspace{0.3in} y(x_0)=y_0, \hspace{0.3in} y(x)\in \mathbb{R}^{m}.
\end{equation}
Here, $y(x)$ is the exact solution and RK methods provide an approximation at time $x_{n}=n\,h$, where $h$ is the step-size and $n = 0,1,\cdots$. The general form of RK methods is given by
\begin{align}\label{RK_general_form}
K_{i} & = y_{n-1} + \displaystyle\sum_{j=1}^{s}a_{ij}\, h\, f(K_{j}), \hspace{0.2in} i = 1,2,\cdots,s, \\ 
y_{n} &= y_{n-1} + \displaystyle\sum_{i=1}^{s}b_{i} h f(K_{i}),
\end{align}
where $K_{i}$ are $s$ stage values and $y_n$ is the output value, which is an approximation of the actual solution $y(x_n)$. RK methods are generally represented by a Butcher tableau
\beq
\begin{array}{c|cccc}
c_1 & a_{11} & a_{12} & \cdots & a_{1s}\\
c_2 & a_{21} & a_{22} & \cdots & a_{2s} \\
\vdots & \vdots & \vdots & \ddots & \vdots \\
c_s & a_{s1} & a_{s2} & \cdots & a_{ss} \\
\hline
& b_1 & b_2 & \cdots & b_s 
\end{array},
\eeq
where $b_{i}$ are the quadrature weights and the consistency conditions
\beq
c_{i} = \displaystyle \sum_{j=1}^{s} a_{ij}, \hspace{0.2in}  i=1,\cdots,s,
\eeq
are the abscissas of the method at which the stages $K_{i}$ are evaluated. The RK methods can be divided into two types, explicit and implicit. For explicit RK methods, we have $a_{ij} = 0$, whenever $i \leq j $. This means that the stages $K_{i}$ can be computed sequentially which requires less computational time and hence are favourite for solving ODEs. However, explicit methods are less preferable due to their limitations in stability for solving stiff differential systems and the inability to preserve quadratic invariants of conservative differential equations.

RK methods are implicit if $a_{ij} \neq 0$ whenever $i \leq j $. In order to solve an $m$ dimensional system of ODEs, for an $s$ stage implicit RK method, $s\,m$ non-linear equations representing the stages need to be solved. This is usually achieved by modified Newton iterations, which are expensive. Hence, the general implicit RK methods are at a disadvantage compared to their explicit counterpart when considering the cost of implementation. However, the advantages of implicit RK methods are in the use for solving stiff differential equations as well as Hamiltonian and structure-preserving ODEs \cite{SRK_Sanz}. The most famous implicit RK methods are the Gauss-Legendre RK methods which are based on shifted Legendre polynomials such that the abscissa $c_i$ of the RK methods are the zeros of the shifted Legendre polynomials $P_{s}^{*}$ on the interval $[0,1]$, given by
\bea
P_{s}^{*}(x)=\frac{s!}{2s}\displaystyle \sum_{k=0}^{s}(-1)^{s-k}\left(\begin{array}{c}
s\\
k 
\end{array}\right)\left(\begin{array}{c}
s+k\\
k 
\end{array}\right)x^{k}.
\eea
An example of one stage, implicit RK method of order two is the implicit mid-point rule
\beq
\renewcommand{\arraystretch}{1.5}
\begin{array}{c|c}
\tfrac{1}{2}&\tfrac{1}{2}\\
\hline
 & 1 
\end{array}.
\eeq
For $s=2$, we have two stages, and the Gauss RK method of order four is given by
\begin{equation}\label{Gauss24}\renewcommand{\arraystretch}{1.5}
\begin{array}{c|cc}
\tfrac{1}{2}-\tfrac{\sqrt{3}}{6} & \tfrac{1}{4}&\tfrac{1}{4}-\tfrac{\sqrt{3}}{6}\\
\tfrac{1}{2}+\tfrac{\sqrt{3}}{6} &\tfrac{1}{4}+\tfrac{\sqrt{3}}{6}&\tfrac{1}{4}\\
\hline
 & \tfrac{1}{2} & \tfrac{1}{2}
\end{array}.
\end{equation}
The values of the coefficients $b_{i}$ and $a_{ij}$ are calculated from the abscissa $c_{i}$ in a way to ensure that the order of the method is $2s$. 

Symplectincess is a characterization of the Hamiltonian systems in terms of their solutions, rather than in terms of the actual form of differential equations. The RK methods which preserve various quantities including the symplectic structure of the solutions of Hamiltonian systems are usually known as symplectic RK integrators and satisfy the symplectic condition \cite{SRK_Sanz,Suris_SRK,Stability_cooper,SRK_LAS}
\beq\label{symp-cond}
b_{i}a_{ij}+b_{j}a_{ji}-b_{i}b_{j}=0, \quad  i,j=1,\cdots,s.
\eeq
All implicit Gauss type RK methods including (\ref{Gauss24}) satisfy the symplectic condition (\ref{symp-cond}) and are well-suited for long-time integration of Hamiltonian systems. However, explicit RK methods do not satisfy the symplectic condition (\ref{symp-cond}), thus they are not symplectic methods. Moreover, they are not suitable for long time integration of Hamiltonian systems because they introduce non-Hamiltonian perturbations which throw the solution out of the Hamiltonian regime.

The advantage of using the symplectic RK method for solving the Hamiltonian systems is that the symplectic methods preserve the quadratic first integrals including the Hamiltonian numerically, while non-symplectic methods do not. One can also use partitioned RK methods which act as a symplectic explicit integrator, but only for separable Hamiltonian differential equations. However, in this paper, we are considering non-separable Hamiltonian differential equations, thus we will not be discussing symplectic explicit partitioned RK methods. For further details, see \cite{SRK_Sanz}.    

\subsubsection{G-symplectic GLMs}

The GLMs are the multi-stage and multi-value numerical algorithms, generalization of RK as well as multi-step methods, used to approximate the numerical solution of a system of ODEs. The general form of GLMs is given by \cite{Butcher:2006:GLM:}
\bea\label{GL method}
   Y &=& h(A\otimes I)f(Y)+(U\otimes I)y^{[n-1]},\\
    y^{[n]}&=&h(B\otimes I)f(Y)+(V\otimes I)y^{[n-1]},
\eea
where $f(Y)$ are the stage derivatives corresponding to $s$-stages $Y\in ({\mathbb{R}}^N)^s$, $A\otimes I$ denotes the Kronecker product of matrix $A$ and identity matrix $I$. At the beginning of a step, the initial values are provided to the vector $y^{[n-1]}$ having $r$-components, resulting in an output vector $y^{[n]}$, can be written in the form
\bea
Y &=& \begin{bmatrix}
Y_{1} \\
Y_{2} \\
\vdots \\
Y_{s}
\end{bmatrix},
  \,\,\,\,\
f(Y) = \begin{bmatrix}
f(Y_1) \\
f(Y_2) \\
 \vdots \\ f(Y_s) 
\end{bmatrix},\\
  y^{[n-1]} &=& \begin{bmatrix}
           {y_1}^{[n-1]} \\
           {y_2}^{[n-1]} \\
           \vdots \\
           {y_r}^{[n-1]}
         \end{bmatrix},
          \,\,\,\,\
  y^{[n]} = \begin{bmatrix}
           {y_1}^{[n]} \\
           {y_2}^{[n]} \\
           \vdots \\
           {y_r}^{[n]}
         \end{bmatrix}.
\eea
The GLMs can also be written in the form
\bea
    Y&=hAf(Y)+Uy^{[n-1]},\\
    y^{[n]}&=hBf(Y)+Vy^{[n-1]},
\eea
where the matrices $A, B, U$, and $V$ usually given as
\beq
M=\left[ {\begin{array}{c|c}
 A & U \\ \hline
B & V
 \end{array} } \right].
\eeq
For example, one-step RK method $[A, b^t, c]$ with one input value $(r=1)$ can be written as a GLM in the form 
\begin{equation}\label{matrix_GL method}
\left[ {\begin{array}{c|c}
 Y\\ \hline
y^{[n]} 
 \end{array} } \right]=\left[ {\begin{array}{c|c}
 A & \mathbf{1} \\ \hline
b^t & 1
 \end{array} } \right]
 \left[ {\begin{array}{c|c}
 hf(Y)\\ \hline
y^{[n-1]} 
 \end{array}} \right].
\end{equation}
A GLM is preconsistent if there exists a preconsistency vector $q_0\in {\mathbb{R}}^r$ such that $Uq_0=\mathbf{1}, Vq_0=q_0$, where $\mathbf{1}$ is the $s$-dimensional unit vector. However, GLM is consistent if it is preconsistent with $q_0$ and there exists a consistency vector $q_1\in {\mathbb{R}}^r$ such that $B\mathbf{1}+Vq_1=q_0+q_1$. A GLM is zero stable if the matrix $V$ is power bounded. The consistency and zero stability are necessary and sufficient conditions for the convergence of a GLM \cite{jack:zdz:2009:}.

The GLMs are not symplectic due to multi-value in nature, therefore can not preserve true quadratic behaviour,
\beq
\langle y,y \rangle = y^{T} S\, y,
\eeq
where $S$ represents the symmetric matrix. However, the extended canonical behaviour
\begin{equation}\label{G_cond}
\langle y^{[n]},y^{[n]} \rangle_{G}=\langle y^{[n-1]},y^{[n-1]} \rangle_{G},
\end{equation}
could be possible to nearly preserve. Here, $G \in {\mathbb{R}}^{r\times r}$ is a symmetric matrix and
\beq
\langle y,z \rangle_{G}=\displaystyle\sum_{i,j=1}^{r}k_{ij}\langle y_{i},z_{j} \rangle.
\eeq
The methods satisfying Eq.~(\ref{G_cond}) are named G-symplectic GLMs and fulfil the given algebraic conditions \cite{Butcher:2006:GLM:}
\bea
G &=& V^{T} G V,\\
DU &=& B^{T} G V,\\
D N + N^{T} D &=& B^{T} G B,
\eea
where $D \in {\mathbb{R}}^{s\times s}$ is a diagonal matrix. The G-symplectic GLMs of order four with two stages were proposed by Butcher \cite{Butcher:2006:GLM:}, and they suffer from parasitic corruption \cite{habib:2010:thesis}. Parasitic solutions are the numerical solutions that can be obtained in addition to the numerical approximation of the exact solution. The multi-value methods generally suffer from parasitic solutions, and GLMs are no exception. However, the G-symplectic GLMs face hazards from their parasitic components similar to those encountered by standard linear multi-step methods, due to the fact that perturbation in non-principle components of the numerical solution is extended by the integration process. The backward error analysis becomes usable to show that linear growth in parasitic components could be bounded by establishing $BU=0$. The parasitic-free GLMs have been developed in \cite{Butcher-etal:2014:,Citro:2020:}.

An example of parasitic free G-symplectic GLM of order four with three input values was constructed in \cite{Habib:2010:thesis:} and given as follows,
\bea
A &=&
\begin{bmatrix} 
	0 & 0 & 0 & 0 \\\\
	 -\frac{11}{127} & \frac{1}{4} & 0 & 0 \\\\
	-\frac{2647}{72240} & \frac{1009}{1680} & \frac{1}{4} & 0 & \\\\
 -\frac{169}{1680} & \frac{113821}{283920} & \frac{473}{676} & 0 &
\end{bmatrix},\\\
U &=& \begin{bmatrix} 
	1 & \frac{1}{4} & \frac{\sqrt3}{4}\\\\
	1 & u_{22} & u_{23}\\\\
	1 &  -u_{22} & -u_{23} \\\\
    1 & -\frac{1}{4} & -\frac{\sqrt3}{4}
\end{bmatrix},\\\
B &=& \begin{bmatrix} 
    -\frac{169}{3360} & \frac{1849}{3360} & \frac{1849}{3360} & -\frac{169}{3360}\\\\
	-\frac{169}{1680} & -\frac{84839}{283920} & \frac{84839}{283920} & \frac{169}{1680} \\\\
	 0 & -\frac{43\sqrt{14595}}{35490} & \frac{43\sqrt{14595}}{35490} & 0
\end{bmatrix},\\\
V &=& \begin{bmatrix} 
	1 & 0 & 0\\\\
	0 & -\frac{1}{2} & -\frac{\sqrt3}{2}\\\\
	0 & \frac{\sqrt3}{2} & -\frac{1}{2}
\end{bmatrix},
\eea
where
\bea
u_{22} &=&-\frac{1973}{29068}+\frac{2\sqrt3\sqrt{14595}}{7267},\\\
u_{23} &=& -\frac{1973\sqrt3}{29068}-\frac{2\sqrt{14595}}{7267}.
\eea
The G and D matrices take the form
\bea
G &=& \begin{bmatrix} 
	1 & 0 & 0\\\
	0 & -\frac{1}{4} & 0\\\
	0 &  0 & - \frac{1}{4}
\end{bmatrix},\\\
D &=& \begin{bmatrix} 
	-\frac{169}{3360} & 0 & 0 &0\\\
	0 & \frac{1849}{3360} & 0 & 0 \\\
	0 &  0 & \frac{1849}{3360} & 0\\\
	0 &  0 & 0 &
 -\frac{169}{3360}
\end{bmatrix}.
\eea
In order to exploit the low computational cost of explicit integrators, we use projection techniques together with explicit schemes for the conservation of quadratic invariants of the underlying differential equations, and this is explored in the following sections.

\subsection{Explicit schemes with projection}

Projection schemes are a standard approach for numerical integration of ODEs on manifolds \cite{Hairer:2000:}. Constructing numerical integrators on manifolds with complex structures is complicated and, therefore, often avoided by embedding the manifold into a larger space with a simple, usually Euclidean structure, where standard integrators can be applied. Projection methods are used to ensure that the solution stays on the correct subspace of the extended solution space, as that is usually not guaranteed by the numerical integrator itself. The idea is to solve the ODE with any numerical integrator and then project the numerical solution onto the desired manifold where the actual solution lies. There are different projection techniques, i.e., standard, symmetric, symplectic, and midpoint projections \cite{kraus:2017:arXiv}. Here, we use a standard projection technique, where projection is applied after each step of the numerical algorithm. In this technique, it is assumed that the solution at the initial conditions lies on the manifold, thus the solution of the projected integrator will also stay on the manifold. This leads to very good long-time stability and improved energy behaviour \cite{Harier:2000:Book,Hairer:2001:,Hairer:2005}.
Consider an IVP 
\beq\label{ODE}
    y'(x)=f(y(x)), \quad y(x_{0}) = y_{0},
\eeq
on a manifold $\mathcal{M}$, and suppose that $y_{n} \in \mathcal{M}$. One-step for standard projection technique $y_{n} \mapsto y_{n+1}$ proceeds as follows
\begin{itemize}
\item compute $\tilde{y}_{n+1} = \psi(y_{n})$, where $\psi$ represents an arbitrary one-step numerical integrator applied to $y'=f(y)$.
\item project the value $\tilde{y}_{n+1}$ onto the manifold $\mathcal{M}$ to obtain $y_{n+1} \in \mathcal{M}$.
\end{itemize}
In the following, we describe the classical RK method of order four with a standard projection technique to preserve the invariants numerically by projecting the solution onto the desired manifold.

\subsubsection{Explicit RK methods with standard projection}

The classical explicit RK method of order four with four stages is given by a Butcher table
\beq\label{Butcher-Tab}
\begin{array}{c|cccc}
0 & & & &\\
\tfrac{1}{2}& \tfrac{1}{2} & & & \\
\tfrac{1}{2}&0& \tfrac{1}{2} & & \\
1& 0 &0 &1 & \\
\hline
& \tfrac{1}{6} &\tfrac{1}{3} &\tfrac{1}{3} & \tfrac{1}{6}  
\end{array}.
\eeq
We suppose that the actual solution $y$ of an IVP remains on an invariant manifold $\mathrm{M}_{y_{0}}$, determined by a known conserved quantity $k(y)$, given by
\beq\label{manifold}
\mathrm{M}_{y_{0}}:= \{y~: ~  k(y) - k(y_{0})=0 \}.
\eeq
In order to solve the system (\ref{ODE}) using explicit RK method (\ref{Butcher-Tab})
with \eqref{manifold}, the numerical solution is desired to stay on the manifold  (\ref{manifold}) that can be achieved with the help of the projection technique whose algorithm for a single step of explicit RK method (\ref{Butcher-Tab}) is as follows
\begin{itemize}
\item For an input vector $y_{n}$, the explicit RK method gives an output vector $\tilde{y}_{n+1}$ that does not remain on the manifold (\ref{manifold}).\\
\item Project $\tilde{y}_{n+1}$ on the manifold (\ref{manifold}) as follows 
\bea\label{maineq}
y_{n+1} &=& \tilde{y}_{n+1}+ \lambda \nabla k(\tilde{y}_{n+1}),\\
\lambda &=& \frac{k(y_0) -k(\tilde{y}_{n+1})}{<\nabla k(\tilde{y}_{n+1}), \nabla k(\tilde{y}_{n+1})>},
\eea
where $\nabla k(y)$ denotes the gradient of $k(y)$. For multiple invariants of an ODE, $k(y)$ is a column vector, and  $\nabla k(y) $ represents the corresponding Jacobin matrix. It is interesting to note that $\nabla k(y) $ is calculated at $\tilde{y}_{n+1}$ rather than $y_{n+1}$ to save the computational cost.
\end{itemize}
The above-mentioned procedure is identical to solving the minimization problem
\beq
\mbox{min} ~ ||y_{n+1}-\tilde{y}_{n+1} ||  \quad \mbox{subject to} \quad y \in \mathrm{M}_{y_{0}},
\eeq
with the standard projection technique to get (\ref{maineq}), where $\lambda$ behaves as a Lagrange multiplier \cite{Hairer:2005,Harier:2000:Book}.

\subsubsection{Explicit GLMs with standard projection}

We can also apply the explicit GLM with the standard projection technique to solve the ODE (\ref{ODE}) subject to the constraint (\ref{manifold}). For this purpose, we consider an explicit GLM of order four, whose matrices A, B, U, and V are given by    
\begin{equation}
\left[
  \begin{array}{c|c}
    A & U \\ \hline
    B & V \\
  \end{array}
\right]=
\left[  \begin{array}{ccccccc}
          0 & 0 & 0 & 0 &\vline 1 &\frac{1}{3}\\
          1 & 0 & 0& 0 &\vline 1 & -\frac{1}{3}\\
          -1 & 1 & 0& 0 &\vline 1 & 1\\
          \frac{3}{8} & \frac{3}{8} & \frac{1}{8} & 0 &\vline 1 & \frac{1}{8}\\ 
          
            \hline
            \frac{3}{8} & \frac{3}{8} & \frac{1}{8} & 0 &\vline 1 & \frac{1}{8}\\
           0 & 0 & 0 & 1 &\vline 0 & 0 \\
         \end{array}
        \right].
\end{equation}
The projection technique will take the form as follows
\begin{itemize}
\item For an input vector $y^{[m]}$, the GLM yields an output vector $\tilde{y}^{[m+1]}$ which does not reside on the manifold (\ref{manifold}). 
\item Pick the first component of the output vector $\tilde{y}_{1}^{[m+1]}$ and  project it onto the invariant manifold (\ref{manifold}) to obtain $y_{1}^{[m+1]} \in M_{y_{0}}$, such that
        \begin{align}\label{mainEq-GL}
        y_{1}^{[m+1]} &= \tilde{y}_{1}^{[m+1]}+ \lambda \nabla g(\tilde{y}_{1}^{[m+1]}),\\
         \lambda&=\frac{k(y_0)-k(\tilde{y}_{1}^{[m+1]})}{<\nabla k(\tilde{y}_{1}^{[m+1]}), \nabla k(\tilde{y}_{1}^{[m+1]})>},
    \end{align}
where $\nabla k(y)$ is the gradient of $k(y)$. For the case of multiple invariants of an ODE, $k(y)$ becomes a column vector, and  $\nabla k (y)$ is the corresponding Jacobin matrix. It is worth mentioning that $\nabla k(y)$ is evaluated at $\tilde{y}_{1}^{[m+1]}$ instead of $y_{1}^{[m+1]}$ to save computational cost. 
\end{itemize}

\end{document}